\documentclass[preprint,12pt]{elsarticle}

\usepackage{amssymb}
\usepackage{amsmath}
\usepackage{parskip}
\usepackage{multirow}

\usepackage{pdfpages}
\usepackage{graphicx}
\usepackage{graphics}
\usepackage {framed}
\usepackage {amsxtra}
\usepackage {enumerate}
\usepackage{commath}
\usepackage[margin= 0.75 in]{geometry}
\usepackage{float}
\usepackage[caption = false]{subfig}
\journal{ar\textit{X}iv}

\begin{document}

\begin{frontmatter}



\title{Accounting for overdispersion and clustering \\ in binomial data from N-of-1 trials \footnote{This work was completed in February, 2020, before the start of the COVID-19 pandemic in the USA, with only minor text edits made in the current version.}}


\author{Majnu John$^{a,}$$^{b,}$$^{c,}$$^{d,}$\footnote{Corresponding author, address: $350$ Community Drive, Manhasset, NY 11030.
 e-mail: {\sf mjohn5@northwell.edu, majnu.john@hofstra.edu}, Phone: +01\,718\,470\,8221, Fax: +01\,718\,343\,1659}, Heejung Bang$^{e}$, Stephanie Winkelbeiner$^{a,b,d,f}$,
 Philipp Homan$^{a,b,d,f,}$}


\begin{abstract}
N-of-1 trials are patient centered randomized controlled trials. Although the primary goal of N-of-1 trials is to obtain the results for each patient separately, pooling
the results across patients also has relevance. In this paper, we present two analytical strategies to pool the results across N-of-1 trials, when the main outcome for each
patient is a binomial variable. Our first method takes into account the extra-binomial variation, while as the second approach takes into account hierarchical clustering in
addition to overdispersion. We illustrate the methods using real data analysis and compare the methods using simulations.

\end{abstract}

\begin{keyword}
N-of-1 trial, overdispersion, quasi-likelihood, random-effects

\end{keyword}

\end{frontmatter}

\section{Introduction}

N-of-1 trials are multi-crossover, randomized, double-blinded experiments conducted with a single patient$^{1}$. Typically in a N-of-1 trial, two interventions A and B are
compared repeatedly in multiple episodes within a patient. These interventions can be two treatments or a treatment and placebo. The order of the pair (AB or BA) is
randomized within each treatment episode. The total number of episodes can vary from patient to patient. As in conventional randomized controlled trials (RCTs), in N-of-1
trials both physician and patient are blinded to the treatments delivered within each episode. As in randomized controlled crossover trials, patients serve as their own
controls thereby reducing the effect of confounding factors in N-of-1 trials. Like crossover trials, N-of-1 trials are feasible only for comparing treatments for chronic
conditions such as for example arthritis, asthma, hypertension, diabetes and chronic pain. Unlike parallel or crossover RCTs, an N-of-1 trial design is more flexible and
individualized$^{2}$. For example, dose levels could be optimized for patients prior to the beginning of the trial so that patients could start the trial at the dose that is
most appropriate for them. Historically, N-of-1 trials have their origin in the field of psychology and education. Although not as much in fervor among practitioners from
other disciplines, the design has been utilized in clinical medicine. This trend is changing currently as funding agencies such as Patient Centered Outcomes Research
Institute (PCORI) have placed emphasis on determining how well a treatment works for a specific patient.

The statistical analysis of N-of-1 trials data is focused on the individual outcome for each patient. Nonetheless, there are considerable benefits related to obtaining pooled
estimates of efficacy from across patients. Zucker \textit{et. al.}$^{3,4}$ were among the first papers that advocated for combining results from N-of-1 trials, citing mainly
population research benefits. Equally important as recognizing the difference between patients, is understanding commonality among patients, including disease processes and
responses to treatments. As mentioned in (Zucker \textit{et. al.}$^{3}$, p.402), ``this commonality forms the basis of medical knowledge and practice, and derives from
reproducibly observed responses of patients to various treatments. Optimal medical treatment depends on recognizing and appropriately balancing the similarities and
differences between patients and their responses to treatments."

Further benefits of obtaining a pooled estimate of efficacy is that it is a measure of success of the individualization done for the set of N-of-1 trials under consideration.
In practice, N-of-1 trials allow modification of treatment parameters (e.g. dose, length of treatment etc.) based on patient preferences. Hence retention rates among enrolled
patients tend to be much higher in such trials compared to conventional RCTs. The conduct of N-of-1 trials also try to exclude patients who do not prefer the treatment(s)
under consideration or terminate early patients for whom benefits are not clearly seen. In this respect, they tend to reach the ``correct" patients for whom the intervention
was really intended. Thus, for example, if in a conventional fixed design RCT the success rate of a drug A over placebo was 20$\%$, we expect to see the pooled success rate
based on N-of-1 trials to be substantially higher (e.g. may be $>$60$\%$) because of the individualization of the treatment parameters. However, if the two rates are very
similar, it is an indication that the individualization done for the set of N-of-1 trials had no additional benefit. Yet another usefulness of calculating the pooled
estimates is in the context of a subgroup analysis. Even if the trial reached the intended patients, it might still be of interest to compare between demographic factors such
as age and sex. Pooled estimates and individual estimates may be of interest in such a subgroup analysis.

From a statistical analysis perspective, one of the main features of N-of-1  trials data that needs to be modeled is the correlation between measurements from repeated
episodes (i.e. within-patient correlation)$^{5}$. Bayesian methods lend themselves easily in such a context and have been used in the analysis of N-of-1 trials. Frequentist
methods have been considered too. For pooling continuous outcome data from a set of N-of-1 trials, various methods have been considered in the literature. Chen and Chen$^{6}$
presented and compared four non-Bayesian methods (paired t-test, two mixed-effect-models approaches and one meta-analysis type pooling method), while Araujo, Julious and Senn$^{7}$ described an empirical Bayesian approach and Zucker \textit{et. al.}$^{3}$ a hierarchical Bayesian random effects model. In the case of a binary outcome, to the best of our knowledge, the only method for pooling the data across subjects that have been considered in the literature so far, is the parametric Bayesian method presented by Schluter and Ware$^{8}$. In this paper we present two frequentist-based methods for pooling binary data from N-of-1 trials.

The first method is a meta-analysis type pooling. This approach takes into account the overdispersion of binomial data obtained by aggregating the binary data for each
patient. Binomial data with an overdispersion parameter can be modeled using beta-binomial distribution. Pooling odds ratios (ORs) based on beta-binomial distribution has
been done in meta-analysis contexts. Bakbergenuly and Kulinskaya$^{9}$ applied Mantel-Haenszel type approaches, which adapted well in obtaining a pooled estimate of OR based on
beta-binomial distribution. Landsman \textit{et. al.}$^{10}$ used a Generalized Estimating Equations (GEE)$^{11}$ based approach to obtain a pooled estimate for a statistic (i.e.
blinding index) based on overdispersed binomial data. The first method considered in our paper uses a quasi-likelihood estimation framework for binomial data from N-of-1
trials accounting for the overdispersion parameter.

Including the overdispersion parameter in a Bernoulli, Binomial or Poisson model is to overcome the restrictive mean-variance relationship prescribed by these models but not
often exhibited by real data. The hierarchical structure in the data (treatment episodes within each patient in N-of-1 trials) are often accommodated by including random
effects into the model. Molenberghs, Verbeke and co-authors$^{12-15}$ presented a modeling strategy in which both overdispersion and hierarchical structure are taken into
account together, for outcomes with Bernoulli, Binomial or Poisson distribution (parametric exponential family of distributions, in general). Prior to their work, modeling
strategies typically considered incorporating either overdispersion only or clustering only. Our second method follows Molenberghs \textit{et al}'s$^{14}$ ideas and adapt their
strategy for pooling binomial data from N-of-1 trials. The second strategy allows us to compensate for the mean-variance link imposed by the Binomial model and also to model
the hierarchical structure present in the N-of-1 design.

The paper is organized as follows. In section 2 we present the methods with technical details. In section 3 we describe the simulation study conducted to compare the methods
and in section 4 we illustrate the application of the methods to analysis of data from three different sets of N-of-1 trials. We conclude with a detailed discussion in the
last section.

\section{Methods}

   We present two approaches to estimate the pooled proportion. For the first approach, we use a beta-binomial distribution to model the overdispersion ($\rho$) in the data
   and quasi-likelihood estimation to determine the pooled estimate. As part of this process, we also estimate the nuisance parameter ($\rho$). In the second approach, we use
   Molenberghs \textit{et al}'s ideas to take into account both overdispersion and hierarchical clustering at the same time. The details are given below. R codes based on
   both methods are given in the Appendix.

  \subsection{Method 1: Quasi-Likelihood \color{black} based approach, which models overdispersion only}

  We assume that there are a total of $J$ patients, with each patient in an N-of-1 trial. For the $j^{th}$ patient there are $n_{j}$ treatment pair episodes. Each episode the
  patient gets treatment A (e.g. control) and B (e.g. experimental) in a randomized order. $X_{j}$ denotes the number of successes out of $n_{j}$ - `success' could mean
  `experimental better than control', or could be just patient preference as in [8]. So the data is essentially \[ (X_{j}, n_{j}), \;\; j = 1, \ldots, J, \] where we assume
  \[X_{j} \sim \mathrm{Binom}(n_{j}, \psi_{j}), \;\; \psi_{j} \sim \mathrm{Beta}(\alpha, \beta), \] which will lead to \[X_{j} \sim \mathrm{BetaBinom}(n_{j}, \theta, \rho), \;\;
  \mathrm{with}\;\; \theta = \frac{\alpha}{\alpha + \beta} \;\; \mathrm{and}\;\;\rho = \frac{1}{\alpha + \beta + 1}\] based on a standard calculation. Our goal is to estimate
  $\theta$ which is the proportion of the treatment preference or proportion of successes for the new treatment. The quasi-likelihood (QL) based estimate of $\theta$ is \begin{equation} \displaystyle \hat{\theta}_{QL} = \frac{\sum_{j=1}^{J}c_{j}X_{j}}{\sum_{j=1}^{J}c_{j}n_{j}} \;\;\mathrm{with}\;\; c_{j} = [1 + (n_{j}-1)\rho]^{-1}. \end{equation} Please see appendix A1 for details of the
  calculation for eq. (1). \color{black} The variance of $\theta_{QL}$ can  be estimated using the Jackknife method, \[ \mathrm{V}^{jk} =
  \mathrm{V}^{jackknife}_{\hat{\theta}_{QL}} = \sum_{j=1}^{J}(\hat{\theta}^{-j}_{QL} - \hat{\theta}_{QL})^{2}, \;\;  \] where $\hat{\theta}^{-j}_{QL}$ is the QL estimate
  obtained with the $j^{th}$ study deleted. The Jackknife method consistently estimates the variance and its use has been discussed in the literature$^{16,17}$.

  There are several methods available in the literature for estimating intra-cluster correlation (ICC) for binary data (e.g.$^{18,19}$). We list a few of them below. One of
  these methods could be used to estimate $\rho$ in equation (1). In some cases the formulas for the estimator of $\rho$ depends on the value of $\theta$. In such cases, we
  alternate between estimating $\theta$ and $\rho$ until convergence for both is reached. Comparison of these estimation methods for $\rho$ has been done in other contexts
  (e.g. $^{20}$). The number of episodes in each N-of-1 trial, $n_{j}$ (also known as the `cluster size'), is typically very small (e.g. $\leq$ 7) compared to the cluster sizes
  commonly seen in cluster randomized trials. Since N-of-1 data has such unique features, we conducted our own simulations to guide the selection of $\rho$. Details of
  simulation study and the conclusions drawn from it are presented in section 3. Here we present the formulas of the estimators of $\rho$. Closed form equations for the
  variances of most of them were first presented in $^{21}$, also available in the literature $^{20,22}$. We present these formulas in Appendix B.

\noindent \textit{The Analysis of Variance (ANOVA) Estimator}.

In the ANOVA framework ICC is considered as the proportion of the total variance explained by the between-subject variance: \[ \frac{\sigma_{b}^{2}}{\sigma_{b}^{2} +
\sigma_{w}^{2}} \] where $\sigma_{b}^{2}$ and $\sigma_{w}^{2}$ denote between-subject and within-subject variances respectively$^{23}$. The estimator used in the continuous
case could be extended to the binary case as well$^{24}$ and the formula is given by \[ \hat{\rho}_{3} = \hat{\rho}_{aov} = \frac{MS_{b} - MS_{w}}{MS_{b} + (K_{0}-1)MS_{w}}, \]
where $MS_{b}$ and $MS_{w}$ are, respectively, the between-subject and within-subject mean squares from a one-way ANOVA table and where \[\displaystyle K_{0} =
\frac{1}{(J-1)}\left[N - \sum_{j=1}^{J} \frac{n_{j}^{2}}{N} \right] \; \mathrm{with}\; N = \sum_{j=1}^{J} n_{j}. \] Explicit formulas for $MS_{b}$ and $MS_{w}$ are
\[\displaystyle MS_{b} = \frac{1}{(J-1)}\left[\sum_{j=1}^{J} \frac{X_{j}^{2}}{n_{j}} - \frac{1}{N}\left(\sum_{j=1}^{J}X_{j} \right)^{2} \right], \;\;\;\;\;\;\; MS_{w} =
\frac{1}{N-J} \left[\sum_{j=1}^{J}X_{j} - \sum_{j=1}^{J}\frac{X_{j}^{2}}{n_{j}} \right]. \] The formula for the variance of this estimate is given in equation B.1.

\noindent \textit{The Fleiss-Cuzick (FC) Estimator}.

FC estimator is similar to Cohen's kappa$^{25}$ used to assess inter-rater reliability. The formula for FC estimator is \[\hat{\rho}_{4} = \hat{\rho}_{fc} = 1 -
\frac{\sum_{j=1}^{J} X_{j}(n_{j}-X_{j})/n_{j}}{(N-J)\theta(1-\theta)}. \] Equation B.2 in the appendix gives for the formula for the variance of this estimator.

\noindent \textit{The Pearson (P) Estimator}.

This estimator is a weighted generalized version of the classical Pearson correlation coefficient calculated over all possible pairs of observations within subjects$^{26,27}$.
We consider the version which assigns equal weight to every pair of observations. The formula for the corresponding estimator is \[\displaystyle \hat{\rho}_{3} =
\hat{\rho}_{p} = \frac{1}{\hat{\theta}_{p}(1-\hat{\theta}_{p})}\left[\frac{\sum_{j=1}^{J}X_{j}(X_{j} - 1)}{\sum_{j=1}^{J}n_{j}(n_{j} - 1) } - \hat{\theta}_{p}^{2} \right],
\;\mathrm{with}\; \hat{\theta}_{p} = \frac{\sum_{j=1}^{J}(n_{j}-1)X_{j}}{\sum_{j=1}^{J}(n_{j}-1)n_{j}}. \] The formula for the variance of the Pearson estimator is given in
equation B.3.

\noindent \textit{The Extended Quasi-Likelihood (EQL) Estimator for $\rho$}.

Differentiating the extended quasi-likelihood function of Nelder and Pregibon$^{28}$ with respect to $\rho$ leads to the following estimating equations for $\rho$:
\begin{equation} \sum_{j=1}^{J}(n_{j}-1)\left(\frac{\Delta_{j} - \phi_{j}}{\phi_{j}^{2}} \right) = 0, \;\mathrm{with}\; \phi_{j} = 1 + (n_{j}-1)\rho = c_{j}^{-1}.
\end{equation}  Here $\Delta_{j}$ is the binomial deviance function \[ \Delta_{j} = 2\left[X_{j}\log\left(\frac{X_{j}}{n_{j}\theta} \right) + (n_{j} - X_{j})\log
\left(\frac{n_{j} - X_{j}}{n_{j} - n_{j}\theta} \right)\right]. \] Note that EQL is different from QL. QL is typically used for estimating the parameters related to the mean
function, while EQL is typically used for estimating the parameters related to the variance function. QL is suited for mean function estimation, but as pointed out in $^{28}$,
Wedderburn's original formulation required knowing the variance function up to a multiplicative constant$^{29}$; using EQL, this requirement can be relaxed.

In general, $\mathbb{E}(\Delta_{j}) \neq \phi_{j}$ and hence eq. (2) is not an unbiased estimating equation; the estimate obtained, $\hat{\rho}_{ql}$, is inconsistent$^{30}$.
However, it has also been shown in previous literature$^{31}$ that the inconsistency of this estimator, is often offset by high efficiency in finite samples. Another important
point regarding this estimator is a practical limitation, which could also lead to biased estimates. If any of the $X_{j}$'s takes values $0$ or $n_{j}$, then $\Delta_{j}$
will become $-\infty$. To prevent this, in practice, we will have to add a small constant to such $X_{j}$'s.  We used $10^{-8}$ as the constant that was added in all our
simulations. We note that there are better suited, but much more involved, methods than simply adding a constant, for example, using data augmentation. Avoiding the problems
related to zero counts is a well-studied area of statistics, mainly by Greenland and co-authors (see $^{32}$ and references within). However, these better suited methods are
mostly Bayesian in nature. We do not pursue these alternate methods since from a practical perspective it is much simpler to add a constant and also because the focus of this
paper is on frequentist approaches.

Finally, we also note that there is no closed-form solution for the variance of the estimate $\hat{\rho}_{ql}$; however, a jackknife based estimate could be computed.

\noindent \textit{The Pseudo-Likelihood (PL) Estimator for $\rho$}.

An unbiased estimator for $\phi_{j}$ is \[ \chi_{j}^{2} = \frac{(X_{j} - n_{j}\theta)^{2}}{n_{j}\theta(1-\theta)}; \;\; \mathrm{clearly}, \;\mathbb{E}(\chi_{j}^{2}) =
\phi_{j}. \] Using $\chi_{j}^{2}$ instead of $\Delta_{j}$ leads to the pseudo-likelihood estimator $\hat{\rho}_{pl}$. Variance estimate for the PL estimator will also be
based on jackknife methods as there is no closed form formula available in the literature.

\noindent \textit{Method of Moments (MM) Estimator for $\rho$}.

A method of moments (MM) based estimator could be obtained by considering the sum of the squared residuals (SSR) given by \[ \sum_{j=1}^{J} \left(\frac{1}{n_{j}[1 +
(n_{j}-1)\rho]}\left\{ \frac{ (X_{j} - n_{j}\theta)^{2} }{\theta(1-\theta)} \right\} \right), \] and setting it equal to its first moment $\mathbb{E}$(SSR). Note that,
although moment based estimation of this type have been considered in the Generalized Linear Models (GLM) context$^{33,34}$, they cannot be adopted to our case by simply
plugging in. Instead, we explicitly derive the above form for SSR and calculate its expectation in Appendix A2.

\noindent \textit{Other estimators for $\rho$}.

There are several estimating approaches for $\rho$, including approaches based on Generalized Estimating Equations (GEE) and Quadratic Estimating Equations (QEE), that we do
not consider in this paper. GEE approach is based on using a working correlation matrix in the quasi-score function. Although GEE is typically considered in a GLM framework
especially for longitudinal data, it could be used also for estimating ICC (see e.g. $^{10}$). Typically, in practice, statistical software such as PROC GENMOD in SAS is
required to implement GEE, which makes it an extra computational step for the analyst. However, if we consider exchangeable correlation structure for the working correlation
matrix, an ICC estimate with a closed form formula is available; this formula, given e.g. in $^{20}$ is essentially the Pearson correlation among residuals within each N-of-1
subject. GEE approach will be especially useful if we are interested in modeling the effect of covariates on the main parameter $\theta$. In a clinical trial setting, the
primary covariate of interest will be an indicator variable for the treatment arms. In an N-of-1 setting, since $\theta$ itself encodes the differences between the two
treatments (e.g. patient preferences for one treatment, or success rate of one treatment over the other), this modeling aspect of GEE is of less value. However, if there are
other covariates of interest (e.g. gender, to understand the sex differences in $\theta$), then GEE approach could be worth considering.

Another interesting class of estimators that we did not consider are the ones based on QEE$^{35}$. QEE in the context of ICC for binary data, has been well-studied in several
papers $^{36-38}$. Gaussian likelihood, quasi-likelihood and profile likelihood estimating equations could be considered as special cases of QEE (see [38]). In order to utilize
the optimal QEE, knowledge about the skewness and kurtosis parameters in the data is required, ideally. In the absence of this knowledge, Paul [37] suggested using second,
third and fourth moments of the beta-binomial distribution as an alternative. Paul \textit{et al}$^{38}$ determined that empirically, in terms of bias and MSE, the estimators
based on optimal QEE is very close to that of AOV estimators. Hence we did not consider the QEE based approach in this paper.

Note also that other types of MM estimators, than the one that we considered above, can be obtained by considering moments for quantities other than the SSR used in our case.
For example, Ridout et al$^{18}$ mentions a class of estimators proposed in $^{39}$ by considering the moments for the weighted residuals of the observed proportion of successes
in the $j^{th}$ group ($j^{th}$ subject in our case). Overall there are about $>$25 estimators for ICC that could be considered, if we include also the minor variations of
all estimators mentioned in Ridout et al $^{18}$. However, Paul \textit{et al}$^{38}$, narrowed it down to a list of 9 estimators, based on their empirical investigation. The
estimators that we considered are the ones that are widely used in other settings (e.g. for CRTs), and also among the narrowed list mentioned in $^{38}$.

\subsection{Method 2: Random effects based approach, which accounts for both overdispersion and clustering} The above approach takes into account only overdispersion. We now
consider an approach introduced by Molenberghs, Verbeke and colleagues$^{12-15}$, that takes into account both overdispersion and clustering. \color{black}
    In this approach we assume,
   \[ X_{j} \sim \mathrm{Binom}(n_{j}, \theta_{j}), \;\;\mathrm{where}\;\; \theta_{j} = \pi_{j}\kappa_{j}, \;\;\pi_{j} \sim \mathrm{Beta}(\alpha, \beta),\;\mathrm{with\;the\;constraint\;}
   \alpha\beta = 1 \] \[ \kappa_{j} = \frac{e^{\xi_{j}}}{1 + e^{\xi_{j}}},\;\; \xi_{j} \sim \mathrm{N}(a, 1).\] Here $\kappa_{j}$ is the random effect parameter that accommodates correlation
   among repeated measures and some overdispersion, and $\pi_{j}$ accounts for the remaining overdispersion. The joint density for data from the $j^{th}$ trial can be written as
   the product of the conditional density of $X_{j}$ given $\xi_{j}$ (i.e. given $\kappa_{j}$) and the marginal density of $\xi_{j}$: \[ f(x_{j}, \xi_{j}) = f(x_{j}, \xi_{j}; \alpha, a, n_{j}) =
   f(x_{j}/\xi_{j})f(\xi_{j}) = \] \[\displaystyle \left\{\sum_{t=0}^{n_{j}-x_{j}} (-1)^{t}\left(\frac{e^{\xi_{j}}}{1+e^{\xi_{j}}} \right)^{x_{j}+t} \frac{n_{j}!}{x_{j}!t!(n_{j}-x_{j}-t)!}
   \frac{B(z_{j}+t+\alpha, \beta) }{B(\alpha, \beta)}\right\} \left(\frac{1}{\sqrt{2\pi}} \exp \left[-\frac{1}{2}(\xi_{j} - a)^{2} \right] \right), \] where $\beta = 1/\alpha$
   and $B(\alpha, \beta)$ denotes the Beta function with parameters $\alpha$ and $\beta$. Then the marginal density is written as \[ f(x_{j}) = f(x_{j}; \alpha, a, n_{j}) =
   \int_{-\infty}^{\infty} f(x_{j}, \xi_{j})d\xi_{j}, \] where the integration is carried out numerically. Finally, the log-likelihood is given by \[ L(\alpha, a) = \sum_{j=1}^{J}
   \log f(x_{j}; \alpha, a, n_{j}). \] Maximum likelihood estimates $\hat{\alpha}$ and $\hat{a}$ are the values of $\alpha$ and $a$ that maximizes the above log-likelihood
   function. Assuming $\pi_{j}$ and $\kappa_{j}$ are independent, we get $\theta = \mathbb{E}(\theta_{j}) = \mathbb{E}(\pi_{j})\mathbb{E}(\kappa_{j})$. $\mathbb{E}(\pi_{j})$ can be estimated as
   $\frac{\hat{\alpha}}{\hat{\alpha} + \hat{\beta}}$. $\mathbb{E}(\kappa_{j})$ is the expectation of the sigmoid function of a normally distributed random variable for which an
   analytical solution does not exist. However, a good approximation for the CDF function is, $\mathrm{Sigmoid}(x) \approx \Phi(\lambda x)$ where $\Phi$ denotes the
   cumulative distribution function (CDF) of a standard normal distribution. A good choice of $\lambda$ obtained by trial and error (plotting one function on top of the
   other) is 0.588. With this, we get a good approximation for $\mathbb{E}(\kappa)$ to be $\Phi(\frac{0.588\hat{a}}{\sqrt{1 + 0.588^{2}}})$. Thus our final estimate of
   $\theta$ for the second method is given by \[ \left( \frac{\hat{\alpha}}{\hat{\alpha} + \hat{\beta}} \right)\Phi\left(\frac{0.588\hat{a}}{\sqrt{1 + 0.588^{2}}} \right).
   \]

   Finally, we note here that, similar to GEE, another advantage of using method 2 is that it is capable of modeling the effects of covariates on the parameters.

\section{Simulation study}

We conducted a simulation study to assess the performance of the two methods proposed in this paper. In this section we present the details about the design and results from
the study. In real conduct of a set of N-of-1 trials, the number of episodes vary from patient to patient, and we mimicked this in our design by letting each patient have
either 2, 3 or 4 episodes. For the main simulation study, at each iteration, we simulated data from 30 N-of-1 trials with 10 trials having 2 episodes, another 10 having 3
episodes and the remaining 10 having 4 episodes, i.e. J = 30. Within each episode, we assume that there are two treatments (A and B in randomized order) being compared and
binary data indicating the patient preference or a more beneficial outcome for one treatment (A) over the other (B).

The total number of iterations for each simulation study was 1000. For each iteration we fixed $\theta$ and $\rho$ and generated $X_{j}$ data, for $j = 1, \ldots, J(=30)$,
from Beta-Binomial($n_{j}, \theta, \rho$). $X_{j}$ is the number of times the $j^{th}$ patient preferred A over B. Recall that $\theta$ is the overall treatment preference
for A at the patient population level and $\rho$ is the overdispersion parameter in the beta-binomial model. We considered different simulation scenarios with $\theta$ set to
0.25, 0.50 or 0.75 for each scenario. Within each such scenario (i.e. $\theta$ fixed) we considered subsidiary scenarios by varying $\rho$. The four choices for $\rho$ that
we considered are 0.01, 0.05, 0.10 and 0.20. $n_{j} =$ 2, 3 or 4 for the main simulation study based on the number of episodes for the $j^{th}$ patient within each iteration.
To understand the effects of increase in sample size, we also conducted simulations with larger sample sizes (e.g. $n_{j} =$ 3, 4, 5 and $J = 45$), which will be
presented towards the end of this section. First we present the results for simulations with $n_{j} =$ 2, 3, 4 and $J = 30$.

We used the `simPop' function from the R package `dirmult'$^{40}$ to generate the trials data within each simulation-iteration. Note that Beta-Binomial is just a special case
of the Dirichlet-Multinomial distribution. Thus, for example, in order to generate 10 trials with 3 episodes for a scenario with $\theta = 0.25$ and $\rho = 0.10$ we took the
first column obtained using the following R command: \\  \verb"      simPop(J = 10, n = 3, pi = c(0.25, 0.75), theta = 0.10)$data"\\
Note that the `theta' in the simPop command corresponds to $\rho$ in our notation, not to be confused with $\theta$ in our notation.

We first focus on simulations results for method 1. Although the standard errors for the $\rho$ estimates are not of primary interest, correctly evaluating them is of
significance when we consider the mean squared errors of the $\rho$ estimates. So, we first make a note about the differences between the two approaches (-closed form
formulas versus jackknife methods-) mentioned above for estimating these standard errors. In Table 1 below (which is continued in Table B1 in appendix B), columns labelled
$\mathrm{se}_{2}$ and $\mathrm{se}_{3}$ give the standard errors for $\hat{\rho}_{aov}$ obtained via the closed-form expression in eq. (B.1) and jackknife methods,
respectively. Standard error, by definition, is just the standard deviation of the sampling distribution. Hence, in a simulation study, the corresponding standard error could
also be obtained simply by taking the standard deviation of the 1000 $\rho$ estimates; $\mathrm{se}_{1}$ represent the standard errors calculated this way. The sampling
distribution for the $\hat{\rho}_{aov}$'s clearly looked like a normal distribution, based on a visual inspection of the corresponding histogram. Hence, we may safely assume
that $\mathrm{se}_{1}$ is almost same as the true value. In tables 1 and B1, we see that $\mathrm{se}_{2}$ is severely biased, but the bias for $\mathrm{se}_{3}$ is
negligible. The large bias in $\mathrm{se}_{2}$ could be due to an error in the formula. Saha and Wang$^{41}$ mentions about minor typographical errors in the original formulas
presented in [21] and based on personal communication from Zou, they present a slightly different formula. Based on Saha and Wang's revision [41], the only difference that
needs to be made in eq. (B.1) is in the term $(N^{2} - J^{2})/\theta(1-\theta)$ that appears in the coefficient of $\rho^{2}$, which has to be changed to $(N^{2} -
J^{2})/N\theta(1-\theta)$. With this revised formula, though, we see roughly $12.4\%$ of the iterations giving negative values for variance estimates of $\hat{\rho}_{aov}$.
If we ignore these negative values, we get standard error estimates closer to $\mathrm{se}_{1}$ values, on the average. For example, with $\theta = 0.25$ and $\rho = 0.01$,
the mean value of the standard error estimates using the revised formula, across all simulations with a positive variance estimate is 0.1231, which is much better than
0.3297, but still quite different from 0.1071. The closed form expression (eq. B.2) for the variance of $\hat{\rho}_{fc}$ also gave negative values; for example, for $\theta
= 0.25$, the proportion of negative variance estimates among all iterations ranged from 1.3$\%$ when true $\rho$ was 0.20 to 14$\%$ when true $\rho$ was 0.01. Note also that
closed form expressions are not available for the EQL-based and PL-based approaches. Considering all the above-mentioned issues for the empirical values for the closed form
estimates and also in order to have consistency across all five types of $\rho$ estimates considered in section 2.1, we utilized only jackknife based methods for variance
estimation while comparing the mean squared errors of the $\rho$ estimates.

\begin{center}
  \tabcolsep=0.11cm
    \begin{tabular}{ | c | c | c | c | c | }                             \hline\hline
    \multicolumn{5}{|c|}{Table 1. $\hat{\rho}_{aov}$ standard errors} \\ \hline
        \;\;\; True $\theta$\;\;\; & \;\;\; True $\rho$\;\;\; & $\mathrm{se}_{1}$ & $\mathrm{se}_{2}$ & $\mathrm{se}_{3}$  \\ \hline
        \multirow{4}{*}{0.25}
        & 0.01 & 0.1071 & 0.3297 & 0.1071  \\ \cline{2-5}
        & 0.05 & 0.1134 & 0.3676 & 0.1147  \\ \cline{2-5}
        & 0.10 & 0.1215 & 0.4549 & 0.1224  \\ \cline{2-5}
        & 0.20 & 0.1340 & 0.7029 & 0.1358  \\ \hline\hline

    \end{tabular}
\end{center}

The mean squared errors (MSEs) for the five $\rho$ estimates for the various true $\theta \leftrightarrow \rho$ combinations are plotted in figure 1. At each iteration, an
MSE value was obtained by adding the squared bias to the jackknife based variance estimate. Each boxplot in figure summarizes the MSEs across all 1000 iterations. It is clear
from the figure that when $\theta$ = 0.50, the EQL estimates for $\rho$ had really bad performance. Although the differences between EQL estimator and other estimators were
not that dramatic in other scenarios ($\theta$ = 0.25 and 0.75), the median MSE for $\hat{\rho}_{eql}$ was still higher than that for the other estimates. To check whether
this difference in MSE for $\hat{\rho}_{eql}$ is due to bias alone, we present the jackknife based variances for all estimators in Table B.2; we see that the variance of the
EQL estimate was also large compared to that for the other $\rho$ estimates, in all scenarios.

Next we consider the MSEs of $\theta$'s which are plotted in figure 2. Note that since $\theta$ is the main parameter of interest, the MSEs of $\theta$ are of more importance
than the MSEs of $\rho$'s. In figure 2, we see that the MSEs of the $\theta$ estimates based on method 1 are all roughly the same. We also see that the MSEs for the $\theta$
estimate based on method 2 is better than the MSEs for other estimators in certain scenarios, but also worse than the other estimators in certain other scenarios. Since it is
not easy to distinguish between the method-1 based estimators in figure 2, we plot the median MSEs separately in figure B.1 in appendix B. Focusing only the results for
method-1 estimates in figure B.1, for the moment, it is clear that $\theta$ estimation with $\hat{\rho}_{eql}$ outperformed $\theta$ estimation with other $\rho$ estimators;
this is in spite of the fact that the MSE for $\hat{\rho}_{eql}$ was worse than that of other $\rho$ estimators, in all scenarios, as we saw from figure 1. We try to make
sense of this surprising result in the next couple of paragraphs.

It is straightforward to explain the above result if $\hat{\rho}_{EQL}$ had low variance in some scenarios, since, based on the Delta method$^{42}$ it is easy to see that
variance of $\hat{\theta}_{QL}$ is directly proportional to the variance of $\hat{\rho}$. The random variables in $\hat{\theta}_{QL}$ are $X_{1}, \ldots, X_{J}$ and
$\hat{\rho}$, Thus, for the moment, if we consider $\hat{\theta}_{QL}$ as a function of the realization of these random variables, $\hat{\theta}_{QL} = h(\hat{\rho}, x_{1},
\ldots, x_{J})$, then based on the Delta method, \begin{equation} \mathrm{Var}\left(\hat{\theta}_{QL} \right) \approx \left(\frac{\partial h}{\partial \hat{\rho}}
\right)^{2}\mathrm{Var}\left(\hat{\rho} \right) + \sum_{j=1}^{J}\left(\frac{\partial h}{\partial x_{j}} \right)^{2}\mathrm{Var}\left(X_{j} \right) +
\;\mathrm{covariance\;terms}. \end{equation} We also note that $\hat{\theta}_{QL}$ is an unbiased estimate of $\theta$; hence, $\mathrm{MSE}(\hat{\theta}_{QL}) =
\mathrm{Var}(\hat{\theta}_{QL})$.

\newpage

\begin{figure}[H]
\begin{center}
\hspace*{-1cm}
\includegraphics[height=6in,width=7in,angle=0]{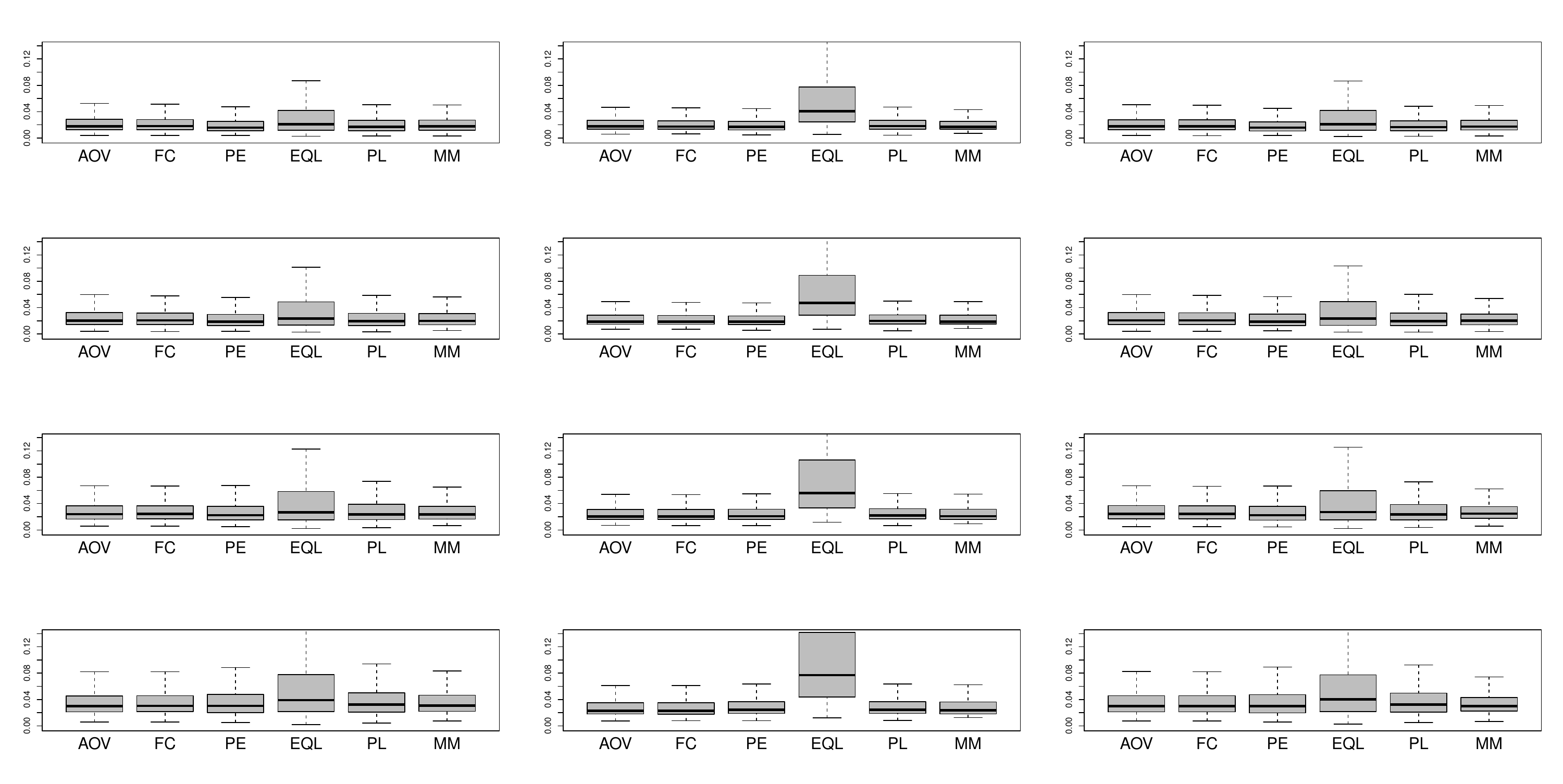}
\caption{Boxplots for the MSEs of $\hat{\rho}$ estimates in various scenarios. Rows, top to bottom, correspond to true $\rho$ values 0.01, 0.05, 0.10 and 0.20. Columns, left
to right, correspond to true $\theta$ values 0.25, 0.50 and 0.75.}
\end{center}
\end{figure}

\begin{figure}[H]
\begin{center}
\hspace*{-1cm}
\includegraphics[height=6in,width=7.2in,angle=0]{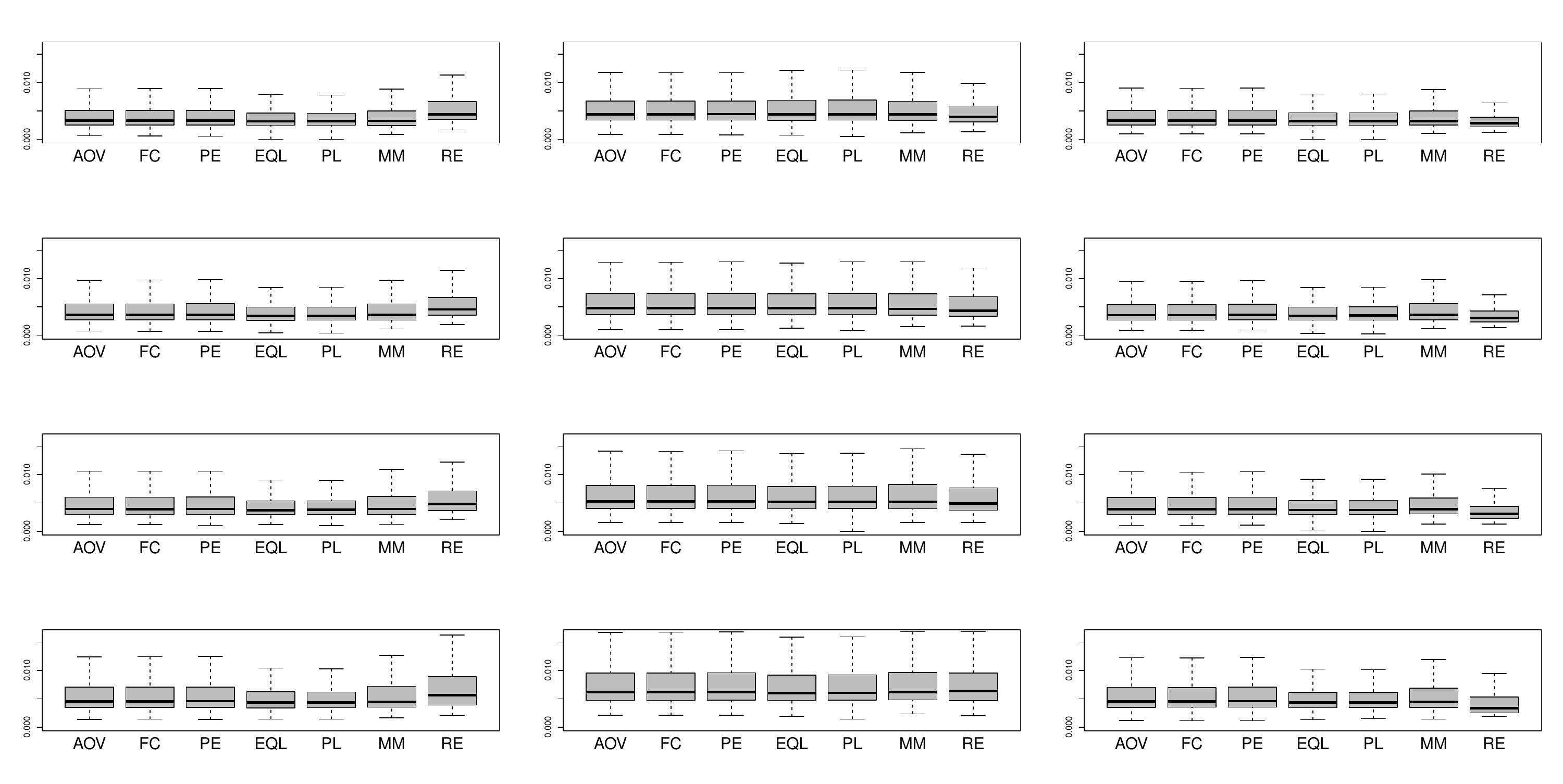}
\caption{Boxplots for the MSEs of $\hat{\theta}$ estimates in various scenarios. Rows, top to bottom, correspond to true $\rho$ values 0.01, 0.05, 0.10 and 0.20. Columns,
left to right, correspond to true $\theta$ values 0.25, 0.50 and 0.75}
\end{center}
\end{figure}

The above considerations would have helped us to understand the reasons for low values of MSE for $\hat{\theta}_{QL}$ if there were certain scenarios where variance of
$\hat{\rho}_{EQL}$ was lower that for the other estimators. However, as we saw in Table B.2 this is the case in none of scenarios. Hence we look a bit more deep and try to
shed some light into the results with further considerations based on the bias in $\hat{\rho}_{EQL}$ observed empirically and behavior of the function $(\partial h/\partial
\hat{\rho})^{2}$. Based on the functional form for $(\partial h/\partial \hat{\rho})^{2}$ given in Appendix C it is clear that it is a function of $\rho$. Since it is a
random function depending on the $X_{j}$'s, it is implicitly a function of $\theta$ too. In figure C.1, we plot it is a function of $\theta$ for fixed $\rho$ values. Each
panel correspond to a fixed $\rho$. For each panel, we considered a grid of $\theta$ values ranging from 0 to 1 with spacing 0.1, and for each grid point we generated 500
data points, $\mathbf{X} = \{X_{1}, \ldots, X_{J}\}$, with $J = 30$ and $n_{j} = 2,3,4$ as before. Based on the data generated at each of the 500 iterations, we calculated
the corresponding $(\partial h/\partial \hat{\rho})^{2}$. Each panel shows the median value of these 500 $(\partial h/\partial \hat{\rho})^{2}$ values, plotted across the
$\theta$ grid. It can be observed from each panel that for certain $\theta$ values, $(\partial h/\partial \hat{\rho})^{2}$ is comparatively peaked, and these peaks vary
across panels (i.e. $\rho$ values). This last observation along with the empirical bias for $\hat{\rho}_{EQL}$ may explain the low MSE for $\hat{\theta}_{QL}$ even in
scenarios where $\hat{\rho}_{EQL}$ has large variance. We elucidate this point with a specific example. Consider the case with true $\rho$ equals 0.10 and true $\theta$
equals 0.50. If $\rho$ is estimated correctly, then the $(\partial h/\partial \hat{\rho})^{2}$ value is comparatively large as seen from the middle panel in figure C.1, which
makes the contribution of the term [$(\partial h/\partial \hat{\rho})^{2} \times \mathrm{Var}(\hat{\rho})$] to $\mathrm{Var}(\hat{\theta}_{QL})$ substantial. However,
$\hat{\rho}_{EQL}$ is typically estimated with bias. For our example, suppose that $\hat{\rho}_{EQL}$ incorrectly estimates $\rho$ to be 0.12. In this case, the $(\partial
h/\partial \hat{\rho})^{2}$ value is almost zero (as seen from the bottom panel), and hence the contribution of the term [$(\partial h/\partial \hat{\rho})^{2} \times
\mathrm{Var}(\hat{\rho})$] almost negligible. Thus, biased estimation could lead to lowering the variance of $\hat{\theta}_{QL}$. Although, this could explain the phenomenon
that we saw for the $\rho$'s that we considered, there is no guarantee that it will be seen for $\rho$'s other than the ones that we considered in our simulation study.

Next we consider the results for $\hat{\theta}$ based on the random effects based approach. Based on figure 2, there were certain scenarios where the MSEs of the random
effects approach were better than those for method 1 based estimation. To better understand the differences between the two overall methods, we also assessed the coverage
probabilities of the 95$\%$ confidence intervals of corresponding $\hat{\theta}$'s based on jackknife standard errors ($\widehat{SE}$). Asymptotic confidence interval for
$\hat{\theta}$ at each simulation iteration can be obtained as ($\hat{\theta} - 1.96\widehat{SE}, \hat{\theta} + 1.96\widehat{SE}$). Coverage for the above confidence
interval was estimated as the percentage of times it contained the true $\theta$ across all 1000 iterations. Estimated coverage for all methods and all scenarios are
presented in table 2.

There was no substantial differences in coverage among all $\theta$-estimates based on method 1. Overall, however, the coverages for $\hat{\theta}$ with EQL and PL approaches
for $\rho$ were similar to each other; also, there were similarities in $\hat{\theta}$ coverages with AOV, FC and PE based approaches for $\rho$ as well. $\hat{\theta}$
coverage with $\hat{\rho}_{EQL}$ and $\hat{\rho}_{PL}$ was consistently the best when true $\rho$ was 0.01, for all $\theta$. When $\theta$ was 0.50, there were two more
$\rho$ values ($\rho$ = 0.05 and 0.10) for which $\hat{\theta}$ with $\hat{\rho}_{EQL}$ and $\hat{\rho}_{PL}$ had the best coverage. In all the remaining cases, EQL and PL
based approaches performed slightly worse than the other approaches. Random effects based approach did fairly well in certain cases (when $\theta$ was 0.25 or 0.50), but did
fairly worse in all cases with $\theta$ = 0.75. This behavior of the random effects approach is not very surprising based on certain facts known about generalized linear
mixed models (GLMMs) in general. To quote verbatim from $^{20}$ (p. 6), ``the sampling distribution of variance estimates in GLMMs is in general strongly asymmetric$^{43,44}$;
even if a standard error is produced by an estimation method, it may be a poor characterization of uncertainty and linear confidence intervals are likely to have poor
coverage properties'', or as mentioned in $^{44}$, ``deviance-based confidence intervals are quite asymmetric, of the form `estimate minus a little, plus a lot'". This applies
not only to variances estimates, but also to main parameter estimates (see e.g. Fig. 1.5 in $^{44}$). We check this is indeed the reason in our case by plotting the histogram of
the $\theta$ estimates obtained by the random effects approach, when true $\theta$ was 0.75 (see figure 3 below); clearly the sampling distributions are skewed in this case.
Skewness is not the case in general; for example, the corresponding histograms in the cases with true $\theta$ 0.25 and 0.50, plotted in figures D.1 and D.2 in Appendix,
clearly shows symmetric sampling distributions. Thus in our case, poor coverage for random effect model based estimates is seen only in the scenarios with asymmetric sampling
distribution. Log transformations as in $^{44}$ may ameliorate the coverage. If bad coverage is an issue in general with random effects approaches which take into account
overdispersion, then variance stabilizing transformations $^{42}$ could be theoretically derived; however, this is beyond the scope of current work.

\begin{center}
  \tabcolsep=0.11cm
    \begin{tabular}{ | c | c | c | c | c | c | c | c | c |}                             \hline\hline
    \multicolumn{9}{|c|}{Table 2. Coverage for $\theta$} \\ \hline
        \;\;\; True $\theta$\;\;\; & \;\;\; True $\rho$\;\;\; & AoV & FC & PE & EQL & PL & MM & RE   \\ \hline
        \multirow{4}{*}{0.25}
        & 0.01 & 94.4$\%$ & 94.4$\%$ & 94.4$\%$ & 94.6$\%$ & 94.3$\%$ & 94.2$\%$ & 94.3$\%$ \\ \cline{2-9}
        & 0.05 & 94.2$\%$ & 94.2$\%$ & 94.0$\%$ & 94.0$\%$ & 93.7$\%$ & 93.9$\%$ & 95.4$\%$ \\ \cline{2-9}
        & 0.10 & 94.0$\%$ & 94.1$\%$ & 94.0$\%$ & 93.6$\%$ & 93.6$\%$ & 94.4$\%$ & 95.6$\%$ \\ \cline{2-9}
        & 0.20 & 93.7$\%$ & 93.7$\%$ & 93.7$\%$ & 93.2$\%$ & 93.2$\%$ & 93.4$\%$ & 95.3$\%$ \\ \hline\hline
        \multirow{4}{*}{0.50}
        & 0.01 & 94.7$\%$ & 94.7$\%$ & 94.7$\%$ & 94.9$\%$ & 94.8$\%$ & 94.5$\%$ & 94.7$\%$ \\ \cline{2-9}
        & 0.05 & 94.3$\%$ & 94.3$\%$ & 94.3$\%$ & 94.8$\%$ & 94.9$\%$ & 95.2$\%$ & 94.2$\%$ \\ \cline{2-9}
        & 0.10 & 94.2$\%$ & 94.2$\%$ & 94.2$\%$ & 95.0$\%$ & 95.6$\%$ & 93.1$\%$ & 93.9$\%$ \\ \cline{2-9}
        & 0.20 & 94.2$\%$ & 94.2$\%$ & 94.3$\%$ & 95.9$\%$ & 95.9$\%$ & 93.0$\%$ & 94.6$\%$ \\ \hline\hline
        \multirow{4}{*}{0.75}
        & 0.01 & 94.1$\%$ & 94.1$\%$ & 94.1$\%$ & 94.2$\%$ & 94.1$\%$ & 93.9$\%$ & 81.8$\%$ \\ \cline{2-9}
        & 0.05 & 94.1$\%$ & 94.2$\%$ & 94.1$\%$ & 93.7$\%$ & 93.4$\%$ & 93.9$\%$ & 82.0$\%$ \\ \cline{2-9}
        & 0.10 & 94.0$\%$ & 94.0$\%$ & 94.0$\%$ & 93.2$\%$ & 93.1$\%$ & 93.4$\%$ & 78.0$\%$ \\ \cline{2-9}
        & 0.20 & 94.0$\%$ & 94.0$\%$ & 93.9$\%$ & 93.3$\%$ & 93.2$\%$ & 94.4$\%$ & 78.3$\%$ \\ \hline\hline

    \end{tabular}
\end{center}

\begin{figure}[H]
\begin{center}
\subfloat{\includegraphics[height=2in, width = 2in]{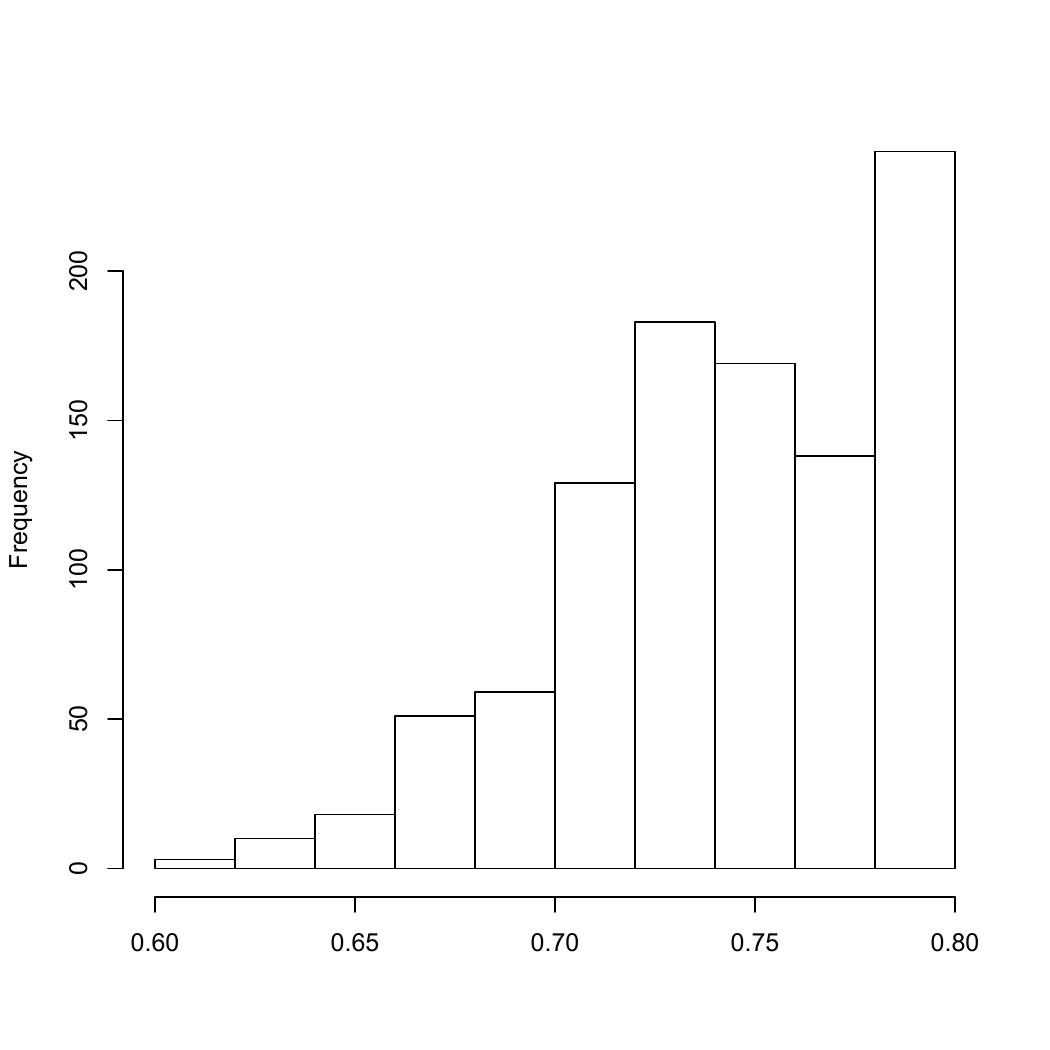}}
\subfloat{\includegraphics[height=2in, width = 2in]{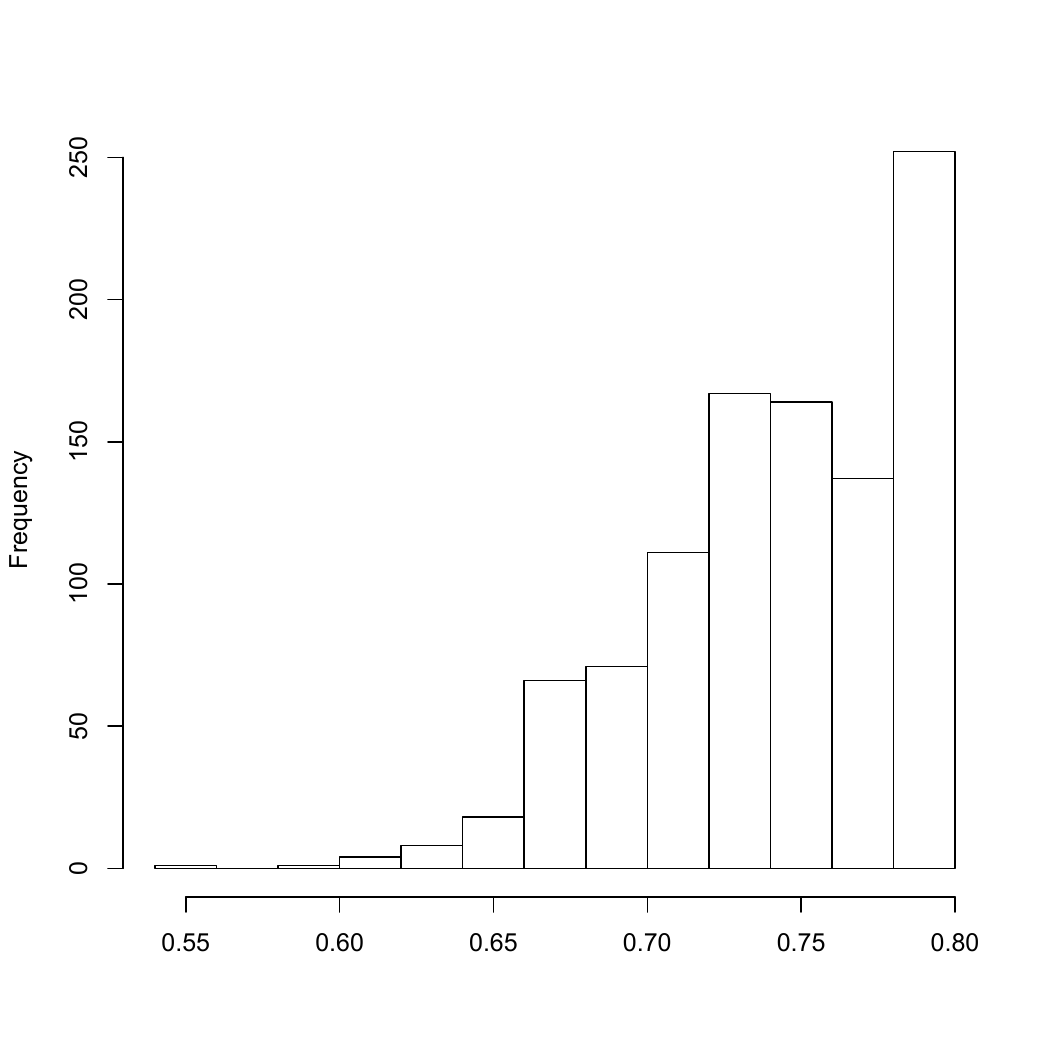}} \\
\subfloat{\includegraphics[height=2in, width = 2in]{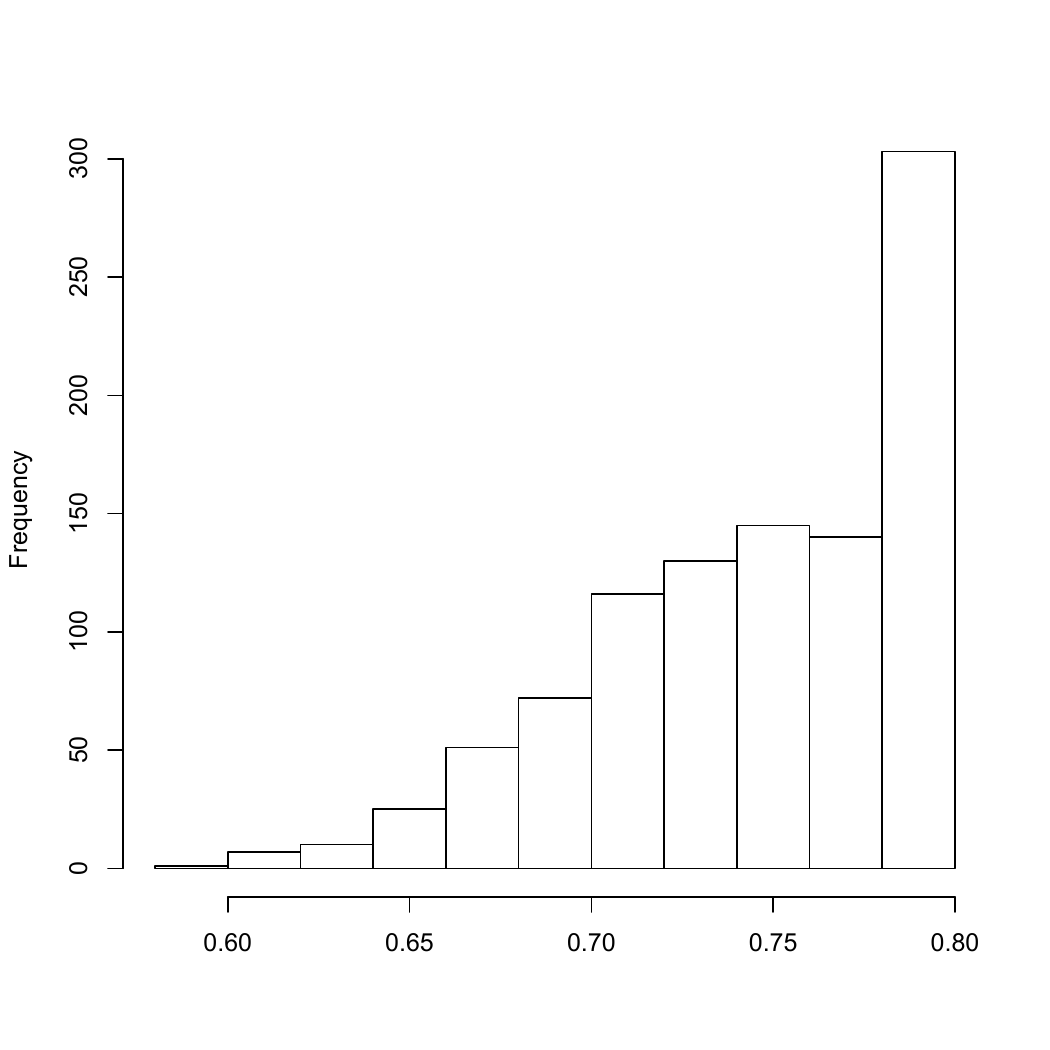}}
\subfloat{\includegraphics[height=2in, width = 2in]{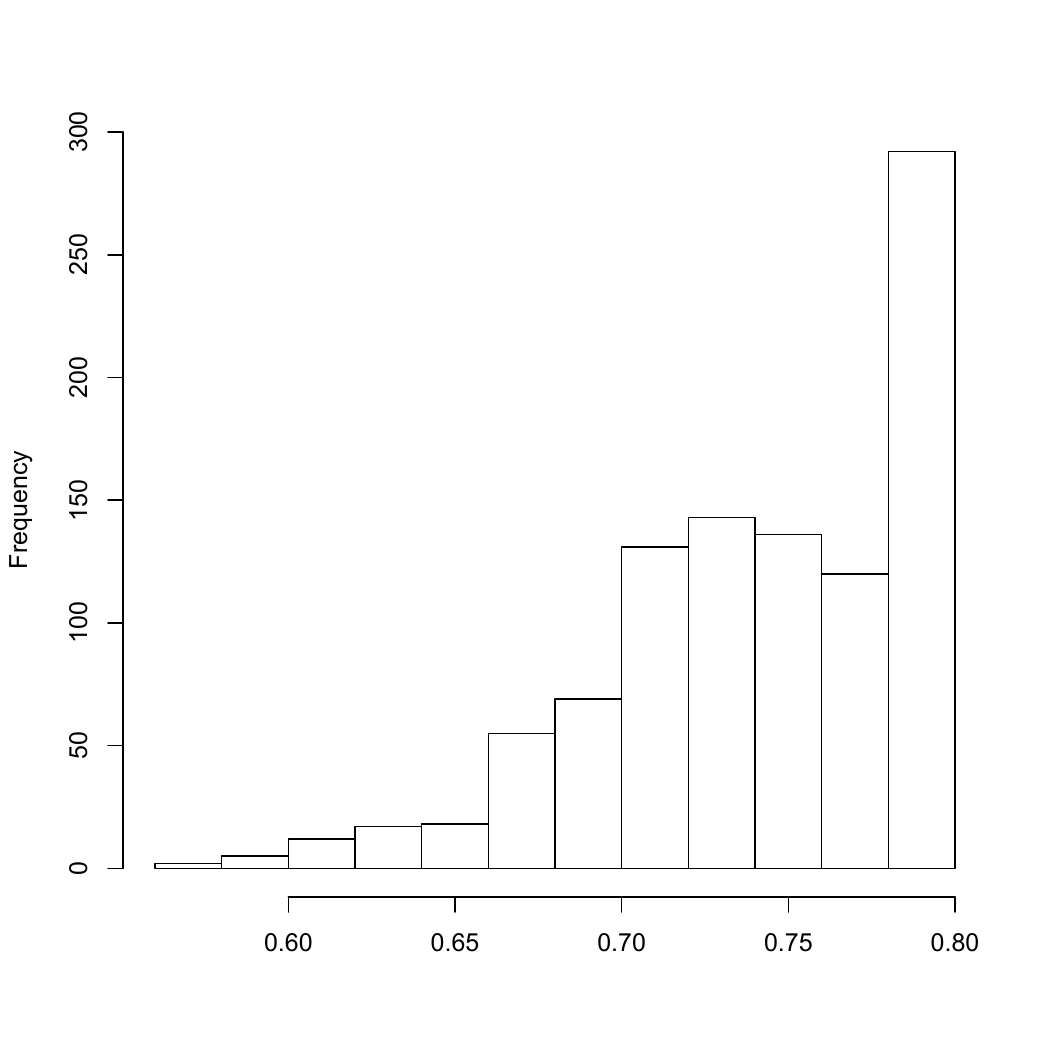}} \\
\caption{Histograms of $\hat{\theta}$ estimates based on method 2, for the cases with true $\theta$ = 0.75}
\end{center}
\end{figure}

Next we consider the effects of increasing the sample size on the $\hat{\theta}$ estimators based on method 1. We did not consider method 2 because of the low coverage seen
above for this method in certain scenarios above. For all the simulations so far the overall number of patients (i.e. $J$) that we considered was 30, split as 10 patients
each with ($n_{j}$ = ) 2, 3 and 4 episodes, respectively. We call this the sample size scenario (a). Sample size can be increased by either increasing $J$ or $n_{j}$ or both.
Thus, we consider three more scenarios corresponding to (b) $J$ = 45 split as 15 patients each with ($n_{j}$ = ) 2, 3 and 4 episodes, respectively; (c) $J$ = 30 split as 10
patients each with ($n_{j}$ = ) 3, 4 and 5 episodes, respectively; and (d) $J$ = 45 split as 15 patients each with ($n_{j}$ = ) 3, 4 and 5 episodes, respectively. Figure 4
shows the boxplots for MSEs for the $\hat{\theta}$'s based on method 1 with various $\rho$ estimates. We present only the scenarios with true $\rho$ = 0.01 and 0.10, and with
$\theta$ = 0.25, as the pattern of results across all scenarios were the same. In all cases, it is obvious that the MSEs get lowered with increase in sample size and the
differences in MSEs between sample size scenarios (a) and (b) are substantial. By considering the middle boxplots within each panel, it is also seen that, the decrease in MSE
is more in scenario (b) (i.e. $n_{j}$'s fixed as in (a), but increased $J$) compared to the decrease in MSE in scenario (c) (i.e. $J$ fixed as in (a), but increased
$n_{j}$'s).

\begin{figure}[H]
\begin{center}
\includegraphics[height=3.5in,width=6in,angle=0]{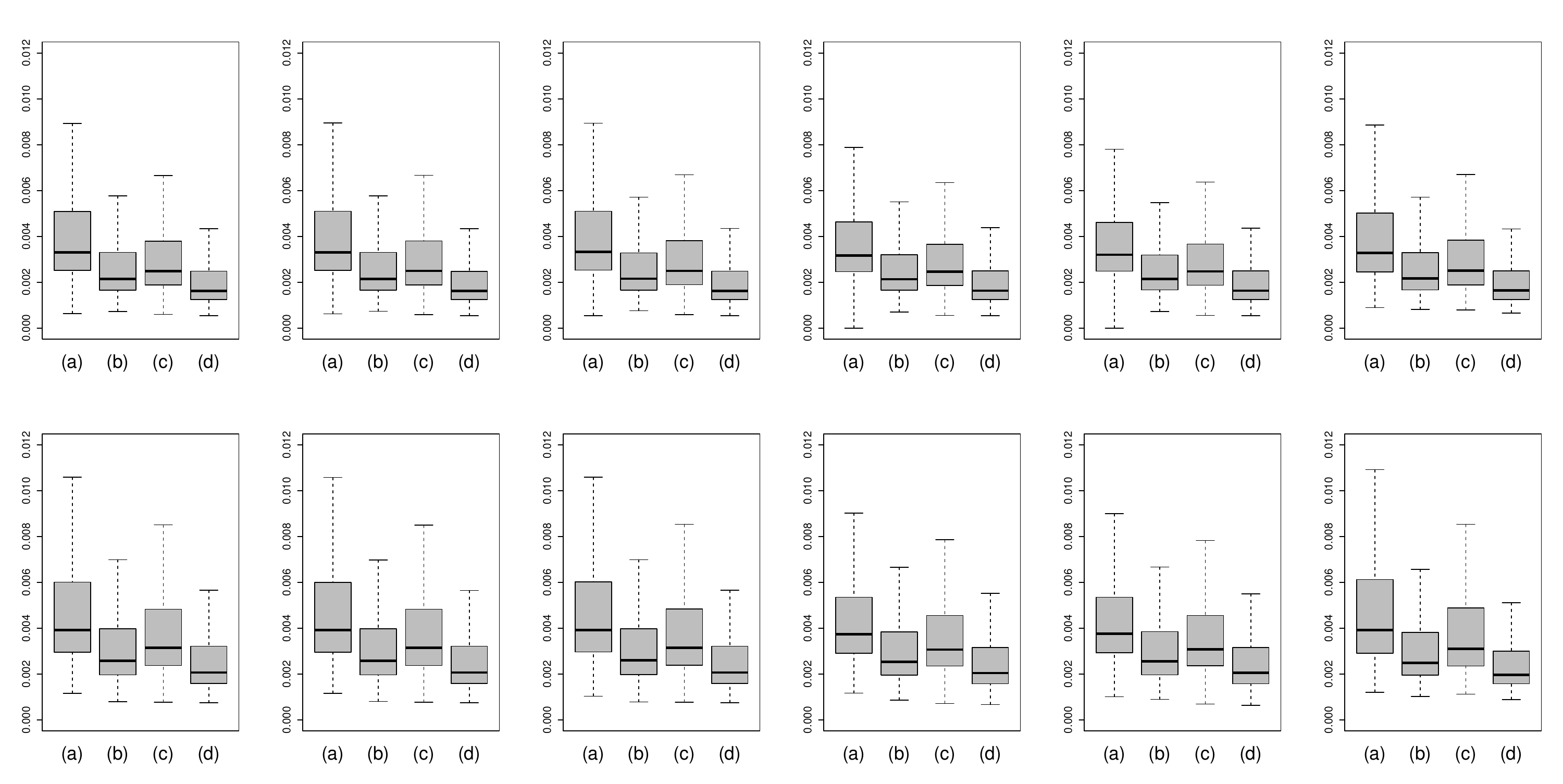}
\caption{Plot to illustrate the effect of sample size increase on estimators based on method 1. Boxplots for $\hat{\theta}$ MSEs for the following sample size scenarios are
plotted within each panel from left to right: (a) J = 30, split as 10 patients each with $n_{j}$ = 2,3,4, respectively; (b) J = 45, split as 15 patients each with $n_{j}$ =
2,3,4 respectively; (c) J = 30, split as 10 patients each with $n_{j}$ = 3,4,5 respectively; (d) J = 45, split as 15 patients each with $n_{j}$ = 3,4,5 respectively. True
$\theta$ in simulations for all panels was 0.25. True $\rho$ was 0.01 for the panels in the top row and 0.10 for the panels in the bottom row. Within each row the panels from
left to right correspond to $\hat{\rho}$ based on AOV, FC, PE, EQL, PL and MM, respectively.}
\end{center}
\end{figure}

\section{Real Data Examples}

We illustrate our methods by applying it to results from three different example sets of N of 1 trials.

\subsection{Amitriptyline for Fibromyalgia}

The first example is based on 23 N-of-1 trials examining the clinical usefulness of amytriptyline for fibromyalgia compared to placebo. The design details of the study and
the data were first presented in $^{45}$. The primary outcome in the original study was the average score from a standardized symptom questionnaire obtained during each
treatment phase. The scores were averaged across all items and the difference in the averaged score between the active drug phase and placebo phase for each patient were
presented in Table 1 in $^{45}$. Schluter and Ware$^{8}$ derived a dichotomous variable for each treatment phase by considering the sign of the above difference-in-average score.
That is, a binary (0-1) variable with 1 indicating that the difference-in-average score is favoring amytriptyline and 0 indicating no superiority of the drug over placebo.
This binary outcome may be interpreted as the treatment preference of each patient at each treatment phase. For the $j^{th}$ patient, based on our previous notation, $n_{j}$
denotes the total number of treatment periods and $X_{j}$ denotes the number of periods for which the patient preferred amytriptyline. The ($X_{j}, n_{j}$) data for the 23
trials used for our analysis are: \begin{align*} & (0,3), (1,3), (2,3), (2,3), (1,3), (2,3), (1,3), (1,3), (1,3), (1,3), (3,4), (4,6), \\ & (2,3), (4,4), (2,3), (3,4), (3,4),
(3,3), (4,4), (3,3), (3,3), (3,3), (3,3). \end{align*} Results are presented in table 3 below.

\begin{center}
  \tabcolsep=0.11cm
    \begin{tabular}{ | c | c | c | }                             \hline\hline
    \multicolumn{3}{|c|}{Table 3. Overall estimates for amytriptyline preference} \\ \hline
      Method & $\hat{\theta}$\;($\widehat{\mathrm{SE}}$) & $\hat{\rho}$\;($\widehat{\mathrm{SE}}$)  \\ \hline\hline
       AOV & 0.6721 (0.0601) & 0.0927 (0.1216) \\ \hline
        FC & 0.6725 (0.0599) & 0.0780 (0.1196) \\ \hline
        PE & 0.6735 (0.0596) & 0.0457 (0.1162) \\ \hline
       EQL & 0.6697 (0.0613) & 0.1904 (0.1705) \\ \hline
        PL & 0.6736 (0.0598) & 0.0439 (0.1392) \\ \hline
        MM & 0.6723 (0.0600) & 0.0828 (0.1257) \\ \hline
        RE & 0.6675 (0.0620) & n/a \\ \hline
    \end{tabular}
\end{center}

There were no substantial differences among the estimates for overall proportion of preference for amytriptyline based on the different approaches mentioned in this paper.
The corresponding jackknife based SEs were also similar. The $\hat{\theta}$ estimates based on method 2 and that based on method 1 with EQL $\rho$-estimation, were slightly
lower, but up to the second decimal place were same as the estimates obtained by other approaches. $\rho$ estimate (0.19) and its SE (0.17) based on EQL approach was very
different from those obtained by other approaches. Based on our simulation study presented in the previous section, we have reasons to believe that these estimates are
biased. Among the other $\rho$-estimation methods, AOV, FC and MM produced similar results. The values based on PE and PL were similar to each other but different from those
obtained by AOV, FC and MM methods. The 95$\%$ confidence intervals for $\rho$ based on all methods contained zero; so, there was no significant evidence of overdispersion
based on any of the methods.

\subsection{NSAIDs vs. Paracetamol for Osteoarthritis}

The next analysis is based on data from a series of N of 1 trials for individual patients with osteoarthritis who have been using non-steroidal anti inflammatory drugs
(NSAIDs) regularly$^{46}$. The main purpose of the trials was to investigate whether paracetamol was as effective as NSAIDs in the treatment of pain and disability related to
osteoarthritis of the hip or knee. Although 13 patients were selected overall, 6 patients did not complete the study. Since our analysis was mainly for illustration, we
included only measures from the 7 patients who completed the study. The primary outcome measures were the total diary score, individual most important complaint and intensity
of pain. Although the three primary outcomes were all continuous or ordinal, binomial data based on these measures were also reported (Table 2, [46]). Binomial data for each
outcome consisted of the number of times the corresponding score for NSAID was better than that for paracetamol among all the treatment periods. The ($X_{j}, n_{j}$) data for
the 7 completed trials for the total diary score are \[ (1,5), (3,5), (2,5), (0,5), (4,5), (3,4), (1,5), \] for the individual most important complaint are \[ (1,5), (0,5),
(2,5), (0,5), (4,5), (2,4), (1,5) \] and for the pain intensity are \[(3,5), (0,5), (2,5), (2,5), (2,5), (3,4), (3,5).\]

The overall $\hat{\theta}$ estimates based on different approaches for each outcome were comparable (see Table 4). EQL and AOV based $\rho$-estimates were different from the
$\rho$ estimates obtained using other methods; EQL based estimate stood out much more than the AOV based estimate. 95$\%$ confidence intervals based on the standard errors
for all $\rho$ estimates contained zero, so that there was no significant evidence of overdispersion.

\begin{center}
  \tabcolsep=0.11cm
    \begin{tabular}{ | l || c | c | c | }                             \hline\hline
    \multicolumn{4}{|c|}{Table 4. Overall estimates for NSAID vs. paracetamol trials} \\ \hline
      Outcome & Method & $\hat{\theta}$\;($\widehat{\mathrm{SE}}$) & $\hat{\rho}$\;($\widehat{\mathrm{SE}}$) \\ \hline\hline
      \multirow{2}{*}{Total daily score}
      & AOV & 0.4166 (0.1236) & 0.1952 (0.1689) \\ \cline{2-4}
      &  FC & 0.4159 (0.1234) & 0.1513 (0.1596) \\ \cline{2-4}
      &  PE & 0.4158 (0.1237) & 0.1488 (0.1642) \\ \cline{2-4}
      & EQL & 0.4170 (0.1234) & 0.2258 (0.2204) \\ \cline{2-4}
      &  PL & 0.4158 (0.1237) & 0.1490 (0.1625) \\ \cline{2-4}
      &  MM & 0.4159 (0.1234) & 0.1518 (0.1582) \\ \cline{2-4}
      &  RE & 0.4175 (0.1152) &  n/a \\ \hline
      \multirow{2}{*}{Individual most important complaint}
      & AOV & 0.2973 (0.1213) & 0.2246 (0.2526) \\ \cline{2-4}
      &  FC & 0.2969 (0.1219) & 0.1794 (0.2391) \\ \cline{2-4}
      &  PE & 0.2970 (0.1224) & 0.1870 (0.2548) \\ \cline{2-4}
      & EQL & 0.2976 (0.1211) & 0.2565 (0.2670) \\ \cline{2-4}
      &  PL & 0.2969 (0.1224) & 0.1788 (0.2459) \\ \cline{2-4}
      &  MM & 0.2968 (0.1219) & 0.1740 (0.2328) \\ \cline{2-4}
      &  RE & 0.2951 (0.1218) &  n/a  \\ \hline
      \multirow{2}{*}{Pain intensity}
      & AOV & 0.4418 (0.0871) &  0.0179 (0.2023) \\ \cline{2-4}
      &  FC & 0.4406 (0.0860) & -0.0140 (0.1787) \\ \cline{2-4}
      &  PE & 0.4404 (0.0857) & -0.0189 (0.1768) \\ \cline{2-4}
      & EQL & 0.4428 (0.0830) &  0.0534 (0.2663) \\ \cline{2-4}
      &  PL & 0.4404 (0.0838) & -0.0204 (0.1895) \\ \cline{2-4}
      &  MM & 0.4406 (0.0849) & -0.0143 (0.1865) \\ \cline{2-4}
      &  RE & 0.4485 (0.0871) &  n/a \\ \hline
    \end{tabular}
\end{center}

One useful aspect of having the overall estimates is in comparing across the outcomes. Overall, NSAID showed better outcomes than paracetamol in only less than 50$\%$
treatment periods. However the percent of treatment pairs in favor of NSAIDs varied based on the outcomes.

\subsection{Theophylline for irreversible CAL} Our third analysis is based on a study$^{47}$ that assessed the rationale for considering theophylline in patients with
irreversible chronic airflow limitation (CAL). The main objective of the study was to determine whether a decision, about theophylline therapy, guided by the more objective
N-of-1 trials framework has any clinically important advantages over a decision based on the standard of practice. Our analysis is focused on the 7 out of 34 patients in the
N-of-1 arm who demonstrated a potentially useful improvement in dyspnea while on theophylline. Dyspnea score was measured on a 7-point Likert scale ranging from 1
(``Extremely short of breath'') to 7 (``Not at all short of breath''). Table 4 in $^{47}$ reported the dyspnea score averaged across the last 5 days of each treatment
(theophylline and placebo) within each period for each of the 7 patients mentioned above. Mahon \textit{et. al.} $^{47}$ suggested that a 0.5 difference in the dyspnea score
between theophylline and placebo to be clinically relevant. We used this threshold to dichotomize the data within each treatment period. Based on this dichotomization, the
($X_{j}, n_{j}$) binomial data for the 7 patients are \[ (0,3), (3,4), (3,3), (2,3), (0,3), (2,4), (2,3)\] where each $X_{j}$ indicates the number of times the dyspnea score
favored theophylline for the $j^{th}$ patient. If we used a stricter criterion of 1-point as the clinically meaningful threshold for the difference in mean dyspnea score,
then the binomial ($X_{j}, n_{j}$) data would be \[ (0,3), (3,4), (3,3), (2,3), (0,3), (1,4), (2,3). \]

\begin{center}
  \tabcolsep=0.11cm
    \begin{tabular}{ | c | c | c || c | c |}                             \hline\hline
    \multicolumn{5}{|c|}{Table 5. Overall estimates of dyspnea score difference favoring theophylline} \\ \hline
      & \multicolumn{2}{|c|}{threshold = 0.5} &  \multicolumn{2}{|c|}{threshold = 1.0} \\ \hline
      Method & $\hat{\theta}$\;($\widehat{\mathrm{SE}}$) & $\hat{\rho}$\;($\widehat{\mathrm{SE}}$) & $\hat{\theta}$\;($\widehat{\mathrm{SE}}$) &
      $\hat{\rho}$\;($\widehat{\mathrm{SE}}$) \\ \hline\hline
       AOV & 0.5159 (0.1499 ) & 0.2836 (0.3021) & 0.4769 (0.1560) & 0.3463 (0.2490) \\ \hline
        FC & 0.5166 (0.1488) & 0.2277 (0.2983) & 0.4770 (0.1551) & 0.2903 (0.2502) \\ \hline
        PE & 0.5174 (0.1482) & 0.1807 (0.2868) & 0.4771 (0.1549) & 0.2582 (0.2224) \\ \hline
       EQL & 0.5144 (0.1516) & 0.4526 (0.4909) & 0.4766 (0.1567) & 0.5220 (0.4048) \\ \hline
        PL & 0.5169 (0.1487) & 0.2064 (0.3393) & 0.4770 (0.1550) & 0.2825 (0.2570) \\ \hline
        MM & 0.5163 (0.1491) & 0.2509 (0.3220) & 0.4770 (0.1552) & 0.3089 (0.2727) \\ \hline
        RE & 0.5075 (0.1540) & n/a & 0.4625 (0.1551) & n/a \\ \hline
    \end{tabular}
\end{center}

Results presented in table 5 show that the $\hat{\theta}$ estimate was slightly lower based on method 2 compared to the estimates obtained with method 1: $\approx 0.01$ and
$\approx 0.015$, respectively, when 0.5 and 1.0 thresholds were used for dypnea score difference. AOV and EQL based $\rho$ estimates differed from the other $\rho$ estimates;
EQL estimate substantially more than the AOV estimate. Based on the wide standard errors for all $\hat{\rho}$'s, there was no statistically significant evidence of
overdispersion, for this dataset as well.

\section{Discussion and Conclusions}

N-of-1 trials are multiple crossover trials conducted in a single patient. The N-of-1 design has several nice features such as randomization, blinding and balance, often
utilized in conventional RCTs. Unlike conventional RCTs, however, N-of-1 trials are more patient-centered. Indeed patient engagement and consideration of patient preferences
are key aspects of the N-of-1 design. In this respect, the design mimics `therapeutic trials' which are utilized extensively in clinical practice but unlike such informal
trials, it provides a sound approach for drawing scientifically valid conclusions. \\

In this paper, we focused on frequentist methods for pooling binomial outcomes from a set of N-of-1 trials. Two aspects of such data that needs to be addressed are
overdispersion and hierarchical clustering effects. We present two approaches in this paper - the first one modeling overdispersion alone and the second one accommodating
both overdispersion and clustering. The first approach used a quasi-likelihood framework while as the second approach was based on random effects modeling framework of
Molenberghs et al$^{12-15}$. Within the first framework we considered several methods available in existing literature for estimating the overdispersion parameter. Jackknife
methods were used to obtain the standard errors for all the parameters involved.

The second approach presented in this paper has a slight edge over the first one in terms of interpretation in a N-of-1 design context. The first method falls under
``population-averaged'' (PA) methods while the second one falls under another big umbrella - ``cluster-specific'' or ``subject-specific'' (SS) methods. In a regression
context (logistic regression specifically or GLM in general) there are several papers (e.g.$^{48-53}$)  that have described in detail the differences in interpretation between
PA and SS approaches. For example, in a logistic regression context for longitudinal data, Hu \textit{et. al.}$^{48}$ mentions ``the PA estimate of time effects does not
distinguish between observations belonging to the same or different subjects.'' However, effects ``estimated from the random-effects models should be interpreted in terms of
the change due to the covariates for a single individual (or, more specifically individuals with the same level on the random subject effect) even if the variable is indeed a
between-subjects factor such as treatment group. Thus, the random-effects model is most useful when inference about individual differences is of major interest.'' In a
logistic regression setting, Neuhaus \textit{et. al.}$^{49}$ compared SS and PA approaches for models that took into account extra-binomial variation (i.e. overdispersion). In
general, SS models are useful when the focus is on making inference at the individual level, whereas PA models are most useful for population-based inference$^{54}$. All the
papers mentioned above compared between PA and SS approaches in a regression setting; however, similar conclusions could be drawn for the methods presented in this paper. In
this regard, since the focus of N-of-1 design is to accommodate individual differences, the interpretation based on the second method might be slightly more appropriate than
that based on the first method.

We conducted simulations and real data analyses to assess the performance of the methods in real practice. Simulations for the first method showed that the mean squared error
for the estimate of the overdispersion parameter to be substantially worse compared to other estimates. However, this did not affect the estimation of the pooled proportion;
mean squared error of the pooled proportion with extended quasi-likelihood estimate for overdispersion was comparable or sometimes even better than the other approaches.
Simulations also showed suboptimal coverage probability for the second approach, in certain scenarios. Real data analyses showed that the pooled proportion based on all the
approaches were very close to each other. Estimates of the overdispersion parameters based on extended quasi-likelihood and analysis of variance differed from those obtained
with other methods.

\section*{Acknowledgements}
Majnu John was  supported  in  part  by  grants  from  the  National Institute  of  Mental  Health  for  an  Advanced  Center  for  Intervention and Services Research (P30
MH090590) and a Center for Intervention Development and Applied Research (P50 MH080173). Heejung Bang was partially supported by the U.S. National Institutes of Health grant
UL1 TR001860.

\section*{Author Affiliations}

$^{a}$ Institute of Behavioral Science, The Feinstein Institute of Medical Research, Northwell Health System, Manhasset, NY, USA. \\
$^{b}$ Division of Psychiatry Research, Zucker Hillside,Northwell Health System,  Glen Oaks, NY, USA. \\
$^{c}$ Department of Mathematics, Hofstra University, Hempstead, NY, USA. \\
$^{d}$ Department of Psychiatry, Zucker School of Medicine at Northwell, Hofstra University, Hempstead, NY, USA. \\
$^{e}$ Division of Biostatistics, Department of Public Health Sciences, University of California, Davis, CA, USA.\\
$^{f}$ Department of Psychiatry, Psychotherapy and Psychosomatics, University of Zurich, 8032, Zurich, Switzerland.\\

\section*{Appendices}

\subsection*{Appendix A1. Calculation of Quasi-likelihood based estimate for $\theta$}
\setcounter{equation}{0}
\renewcommand{\theequation}{A.\arabic{equation}}
\setcounter{figure}{0}
\renewcommand{\thefigure}{A.\arabic{figure}}
Since this is a standard calculation adapted to the N-of-1 setting, we present it in the appendix. For convenience with algebra, we write the data as \[ (X_{1j}, X_{2j},
n_{j}), \;\; j = 1, \ldots, J, \;\; \mathrm{where}\;\; X_{1j} = X_{j},\;\; X_{2j} = n_{j} - X_{j};\;\; \mathrm{denote}\;\; \mathbf{X}_{j} = [X_{1j}, X_{2j}]^{T}, \] \[X_{1j}
\sim \mathrm{BetaBinom}(n_{j}, \theta_{1}, \rho)\;\; X_{2j} \sim \mathrm{BetaBinom}(n_{j}, \theta_{2}, \rho)\;\; \mathrm{where}\;\; \theta_{2} = 1 - \theta_{1}. \] Although
we introduced these new terms, we are interested in only estimating $\theta_{1} (= \theta)$. We consider \[\mathbf{Z}_{j} = [Z_{1j}, Z_{2j}]^{T}, \;\; \mathrm{where}\;\;
Z_{1j} = X_{1j} - n_{j}\theta_{1}, \;\; Z_{2j} = X_{2j} - n_{j}\theta_{2};\;\;\mathrm{note\;\;that}\;\; Z_{1j} + Z_{2j} = 0. \] We first assume that $\rho$ is known. The
Quasi-Likelihood (QL) equations for estimating $\theta$ will be \begin{equation} \sum_{j=1}^{J} \mathbf{D}_{j}\mathbf{S}_{j}^{-1}\mathbf{Z}_{j} = 0. \end{equation} Here
$\mathbf{D}_{j} = \mathrm{diag}(n_{j}, n_{j})$ is the
  matrix of partial derivatives of $\mathbb{E}(\mathbf{X}_{j})$ with respect to $\theta_{1}$ and $\theta_{2}$, and \[ \mathbf{S}_{j} = n_{j}[1 + (n_{j}-1)\rho]\left\{
  \left[ \begin{array}{cc} \theta_{1} & 0 \\
   0 & \theta_{2} \end{array}  \right] - \left[ \begin{array}{c} \theta_{1} \\ \theta_{2} \end{array} \right]\left[\begin{array}{cc} \theta_{1}, \theta_{2} \end{array}
   \right] \right\} = n_{j}[1 + (n_{j}-1)\rho] \left[ \begin{array}{cc} \theta_{1}(1-\theta_{1})  & -\theta_{1}\theta_{2} \\ -\theta_{2}\theta_{1} & \theta_{2}(1-\theta_{2})
   \end{array}  \right]\] Note that determinant of $\mathbf{S}_{j}$ is zero and hence $\mathbf{S}_{j}$ is not invertible. In order to get around this problem we could take a
   generalized inverse [see p.168, in [55]) or we could consider \[ \mathbf{S}_{j, \varepsilon} = n_{j}[1 + (n_{j}-1)\rho]\left\{   \left[ \begin{array}{cc} \theta_{1} & 0 \\
   0 & \theta_{2} \end{array}  \right] - (1- \varepsilon)\left[ \begin{array}{c} \theta_{1} \\ \theta_{2} \end{array} \right]\left[\begin{array}{cc} \theta_{1}, \theta_{2}
   \end{array} \right] \right\} \] with the idea that we will take $\varepsilon \rightarrow 0$ at the end. However, we will, in a moment, note that the term with
   $(1-\varepsilon)$ will disappear in the final calculation, thus essentially reducing $\mathbf{S}_{j, \varepsilon}^{-1}$ to a generalized inverse. \[ \mathbf{S}_{j,
   \varepsilon}^{-1} = \left\{n_{j}[1 + (n_{j}-1)\rho] \right\}^{-1} \left\{\mathrm{diag}(\theta_{1}^{-1}, \theta_{2}^{-1}) - \frac{1-\varepsilon}{\varepsilon}
   \mathbf{J}_{2}\right\},\;\mathbf{J}_{2}\; \mathrm{is\; the}\; 2 \times 2\; \mathrm{matrix\; of\; ones}. \] Since $\mathbf{J}_{2}\mathbf{Z}_{j} = \mathbf{0}$ we get \[
   \mathbf{D}_{j}\mathbf{S}_{j, \varepsilon}^{-1}\mathbf{Z}_{j} = \frac{1}{1 + (n_{j}-1)\rho}\left[ \begin{array}{c} Z_{1j}/\theta_{1} \\ Z_{2j}/\theta_{2} \end{array}
   \right]. \] (Note the absence of $\varepsilon$ in the final term.) Since we are interested only in estimating $\theta_{1}$ we need only the first row from equation (A.1)
   with $\mathbf{S}_{j, \varepsilon}^{-1}$ substituted for $\mathbf{S}_{j}^{-1}$: \[ \sum_{j=1}^{J} \left[ \frac{ \left(X_{1j}/\theta_{1} \right) - n_{j} }{1 + (n_{j}-1)\rho
   } \right] = 0 \] \[ i.e.\;\; \frac{1}{\theta_{1}} \times \sum_{j=1}^{J}c_{j}X_{1j} = \sum_{j=1}^{J}c_{j}n_{j}, \;\;\mathrm{with}\;\; c_{j} = [1 + (n_{j}-1)\rho]^{-1} \]
   from which we get the QL estimate of $\theta_{1} (= \theta)$ as \begin{equation} \displaystyle \hat{\theta}_{QL} =
   \frac{\sum_{j=1}^{J}c_{j}X_{1j}}{\sum_{j=1}^{J}c_{j}n_{j}} = \frac{\sum_{j=1}^{J}c_{j}X_{j}}{\sum_{j=1}^{J}c_{j}n_{j}}. \end{equation}

\subsection*{Appendix A2. SSR for MM estimation and calculation of $\mathbb{E}$(SSR)}

   In order to estimate $\rho$ using method of moments we consider the sum squares of the residuals, \[ \mathrm{SSR} = \sum_{j=1}^{J} \mathbf{Z}_{j}^{T}\mathbf{S}_{j,
   \varepsilon}^{-1}\mathbf{Z}_{j} = \sum_{j=1}^{J} \left(\frac{1}{n_{j}[1 + (n_{j}-1)\rho]}\left\{ \frac{ (X_{1j} - n_{j}\theta_{1})^{2} }{\theta_{1}} + \frac{ (X_{2j} -
   n_{j}\theta_{2})^{2} }{\theta_{2}} \right\} \right) \] Taking expectations, we get \[ \mathbb{E}(\mathrm{SSR}) = \sum_{j=1}^{J} \left(\frac{1}{n_{j}[1 +
   (n_{j}-1)\rho]}\left\{ \frac{ n_{j}[1 + (n_{j}-1)\rho]\theta_{1}(1-\theta_{1}) }{\theta_{1}} + \mathbb{E}\left[\frac{ (X_{2j} - n_{j}\theta_{2})^{2} }{\theta_{2}} \right]
   \right\} \right)\] \begin{equation} = \sum_{j=1}^{J} \left( 1- \theta_{1} + \frac{1}{n_{j}[1 + (n_{j}-1)\rho]}\mathbb{E}\left[\frac{ (X_{2j} - n_{j}\theta_{2})^{2}
   }{\theta_{2}} \right] \right) \end{equation} \[ \mathrm{Now} \;\; (X_{2j} - n_{j}\theta_{2})^{2} = [n_{j} - X_{1j} - n_{j}(1-\theta_{1})]^{2} = [-X_{1j} +
   n_{j}\theta_{1}]^{2}\;\;\mathrm{so\;\;that} \] \[ \mathbb{E}\left[\frac{ (X_{2j} - n_{j}\theta_{2})^{2} }{\theta_{2}} \right] = \frac{ n_{j}[1 +
   (n_{j}-1)\rho]\theta_{1}(1-\theta_{1}) }{\theta_{2}} = n_{j}[1 + (n_{j}-1)\rho]\theta_{1}. \] Putting this in equation (A.3) we get \[ \mathbb{E}(\mathrm{SSR}) =
   \sum_{j=1}^{J} \left( 1- \theta_{1} + \theta_{1} \right) = J. \] An estimate of $\rho$ may be thus obtained from setting \begin{equation} \mathrm{SSR} = J \end{equation}
   where we may re-write SSR as \[ \sum_{j=1}^{J} \left(\frac{1}{n_{j}[1 + (n_{j}-1)\rho]}\left\{ \frac{ (X_{j} - n_{j}\theta)^{2} }{\theta(1-\theta)} \right\} \right). \] We
   could plug-in $\theta_{QL}$ for $\theta$ in equation (A.4) and alternate between estimating $\theta$ via equation (A.2) and $\rho$ via equation (A.4) until convergence is
   obtained. Note that calculations based on generalized inverse (presented in p. 168, $^{55}$) will result in the same formulas as above.

\subsection*{Appendix B. Variance formulas for estimators of $\rho$}
\setcounter{equation}{0}
\renewcommand{\theequation}{B.\arabic{equation}}
\setcounter{figure}{0}
\renewcommand{\thefigure}{B.\arabic{figure}}

\noindent \textit{Variance formula for the ANOVA estimator}.

The variance of $\hat{\rho}_{aov}$ is given by \[ \mathrm{Var}(\hat{\rho}_{aov}) = \frac{[(J-1)K_{0}N(N-J)]^{2}}{\lambda^{4}} \left\{2J + \left(\frac{1}{\theta(1-\theta)} - 6
\right)\sum_{j=1}^{J}\frac{1}{n_{j}} \right. \] \[ \left. \;\;\;\;\;\;\;\;\;\;\;\;\;+ \left[\left(\frac{1}{\theta(1-\theta)} - 6 \right)\sum_{j=1}^{J}\frac{1}{n_{j}} - 2N +
7J - \frac{8J^{2}}{N} - \frac{2J(N-J)}{N\theta(1-\theta)} + \left(\frac{1}{\theta(1-\theta)} - 3 \right)\sum_{j=1}^{J}n_{j}^{2} \right]\rho  \right. \] \[ \left.
\;\;\;\;\;\;\;\;\;\;\;\;\;+ \left[ \frac{N^{2} - J^{2}}{\theta(1-\theta)} - 2N - J + \frac{4J^{2}}{N} + \left(7 - \frac{8J}{N} - \frac{2(N-J)}{N\theta(1-\theta)}
\right)\sum_{j=1}^{J}n_{j}^{2} \right]\rho^{2}   \right. \] \begin{equation} \left. \;\;\;\;\;\;\;\;\;\;\;\;\;+ \left[\left(\frac{1}{\theta(1-\theta)} - 4
\right)\left(\frac{N-J}{N} \right)^{2}(\sum_{j=1}^{J}n_{j}^{2} - N) \right]\rho^{3} \right\}\; , \end{equation} where $\lambda = (N-J)[N - 1 - K_{0}(J-1)]\rho +
N(J-1)(K_{0}-1),$ (Wu \textit{et al}, [20]). In the above formula, for $\rho$ and $\theta$ we substitute their estimates.

\noindent \textit{Variance formula for FC estimator}.

The formula for the variance of FC estimator of $\rho$ is \[ \mathrm{Var}(\hat{\rho}_{fc}) = (1-\rho)\left\{\left(\frac{1}{\theta(1-\theta)} - 6
\right)\frac{n_{j}^{-1}}{(N-J)^{2}} + \left(2N + 4J - \frac{J}{\theta(1-\theta)} \right)\frac{J}{N(N-J)^{2}}  \right. \] \[ \left. \;\;\;\;\;\;\;\;\;\;\;\;\;+
\left[\frac{\sum_{j=1}^{J}n_{j}^{2}}{N^{2}\theta(1-\theta)} - \frac{(3N-2J)(N-2J)\sum_{j=1}^{J}n_{j}^{2}}{N^{2}(N-J)^{2}} - \frac{2N-J}{(N-J)^{2}} \right] \rho    \right. \]
\begin{equation} \left. \;\;\;\;\;\;\;\;\;\;\;\;\;+ \left[\left(4 - \frac{1}{\theta(1-\theta)}\right)\frac{n_{j}^{2} - N}{N^{2}}  \right] \rho^{2} \right\}. \end{equation} In
this formula too, as in the formula for the variances of ANOVA estimator, we substitute the estimated values for $\theta$ and $\rho$.

\noindent \textit{Variance formula for Pearson estimator}.

The formula for the variance of the Pearson estimator is given by \[ \mathrm{Var}(\hat{\rho}_{p}) = \frac{(1-\rho)}{[\sum_{j=1}^{J}n_{j}(n_{j}-1)]^{2}}
\left\{2\sum_{j=1}^{J}n_{j}(n_{j}-1) + \left[\left(\frac{1}{\theta(1-\theta)} - 3 \right)\sum_{j=1}^{J}n_{j}^{2}(n_{j}-1)^{2} \right]\rho  \right. \] \[ \left.
\;\;\;\;\;\;\;\;\;\;\;\;\;+ \left[\left(4 - \frac{1}{\theta(1-\theta)} \right)\sum_{j=1}^{J}n_{j}(n_{j}-1)^{3} \right]\rho^{2} \right\}. \]


\begin{center}
  \tabcolsep=0.11cm
    \begin{tabular}{ | c | c | c | c | c | }                             \hline\hline
    \multicolumn{5}{|c|}{Table B1. $\hat{\rho}_{aov}$ standard errors (contd.)} \\ \hline
        \;\;\; True $\theta$\;\;\; & \;\;\; True $\rho$\;\;\; & $\mathrm{se}_{1}$ & $\mathrm{se}_{2}$ & $\mathrm{se}_{3}$  \\ \hline
        \multirow{4}{*}{0.50}
        & 0.01 & 0.1055 & 0.2851 & 0.1081  \\ \cline{2-5}
        & 0.05 & 0.1085 & 0.3082 & 0.1126  \\ \cline{2-5}
        & 0.10 & 0.1131 & 0.3797 & 0.1174  \\ \cline{2-5}
        & 0.20 & 0.1208 & 0.6110 & 0.1244  \\ \hline\hline
        \multirow{4}{*}{0.75}
        & 0.01 & 0.1059 & 0.3267 & 0.1069  \\ \cline{2-5}
        & 0.05 & 0.1140 & 0.3703 & 0.1142  \\ \cline{2-5}
        & 0.10 & 0.1208 & 0.4554 & 0.1222  \\ \cline{2-5}
        & 0.20 & 0.1340 & 0.7176 & 0.1363  \\ \hline\hline

    \end{tabular}
\end{center}

\begin{center}
  \tabcolsep=0.11cm
    \begin{tabular}{ | c | c | c | c | c | c | c | c | }                             \hline\hline
    \multicolumn{8}{|c|}{Table B.2. Mean of jackknife based $\hat{\rho}$ variances} \\ \hline
        \;\;\; True $\theta$\;\;\; & \;\;\; True $\rho$\;\;\; & AoV & FC & PE & QL & PL & MM   \\ \hline
        \multirow{4}{*}{0.25}
        & 0.01 & 0.0124 & 0.0121 & 0.0112 & 0.0150 & 0.0118 & 0.0120 \\ \cline{2-8}
        & 0.05 & 0.0141 & 0.0138 & 0.0132 & 0.0175 & 0.0138 & 0.0134 \\ \cline{2-8}
        & 0.10 & 0.0160 & 0.0158 & 0.0155 & 0.0207 & 0.0163 & 0.0157 \\ \cline{2-8}
        & 0.20 & 0.0195 & 0.0194 & 0.0198 & 0.0279 & 0.0209 & 0.0198 \\ \hline\hline
        \multirow{4}{*}{0.50}
        & 0.01 & 0.0119 & 0.0116 & 0.0111 & 0.0257 & 0.0118 & 0.0114 \\ \cline{2-8}
        & 0.05 & 0.0129 & 0.0126 & 0.0123 & 0.0282 & 0.0130 & 0.0124 \\ \cline{2-8}
        & 0.10 & 0.0140 & 0.0137 & 0.0137 & 0.0311 & 0.0144 & 0.0138 \\ \cline{2-8}
        & 0.20 & 0.0156 & 0.0155 & 0.0161 & 0.0350 & 0.0165 & 0.0157 \\ \hline\hline
        \multirow{4}{*}{0.75}
        & 0.01 & 0.0123 & 0.0120 & 0.0111 & 0.0150 & 0.0117 & 0.0120 \\ \cline{2-8}
        & 0.05 & 0.0140 & 0.0137 & 0.0130 & 0.0175 & 0.0138 & 0.0138 \\ \cline{2-8}
        & 0.10 & 0.0160 & 0.0157 & 0.0154 & 0.0206 & 0.0162 & 0.0158 \\ \cline{2-8}
        & 0.20 & 0.0196 & 0.0195 & 0.0199 & 0.0281 & 0.0208 & 0.0197 \\ \hline\hline

    \end{tabular}
\end{center}

\begin{figure}[H]
\begin{center}
\hspace*{-1cm}
\includegraphics[height=6in,width=7in,angle=0]{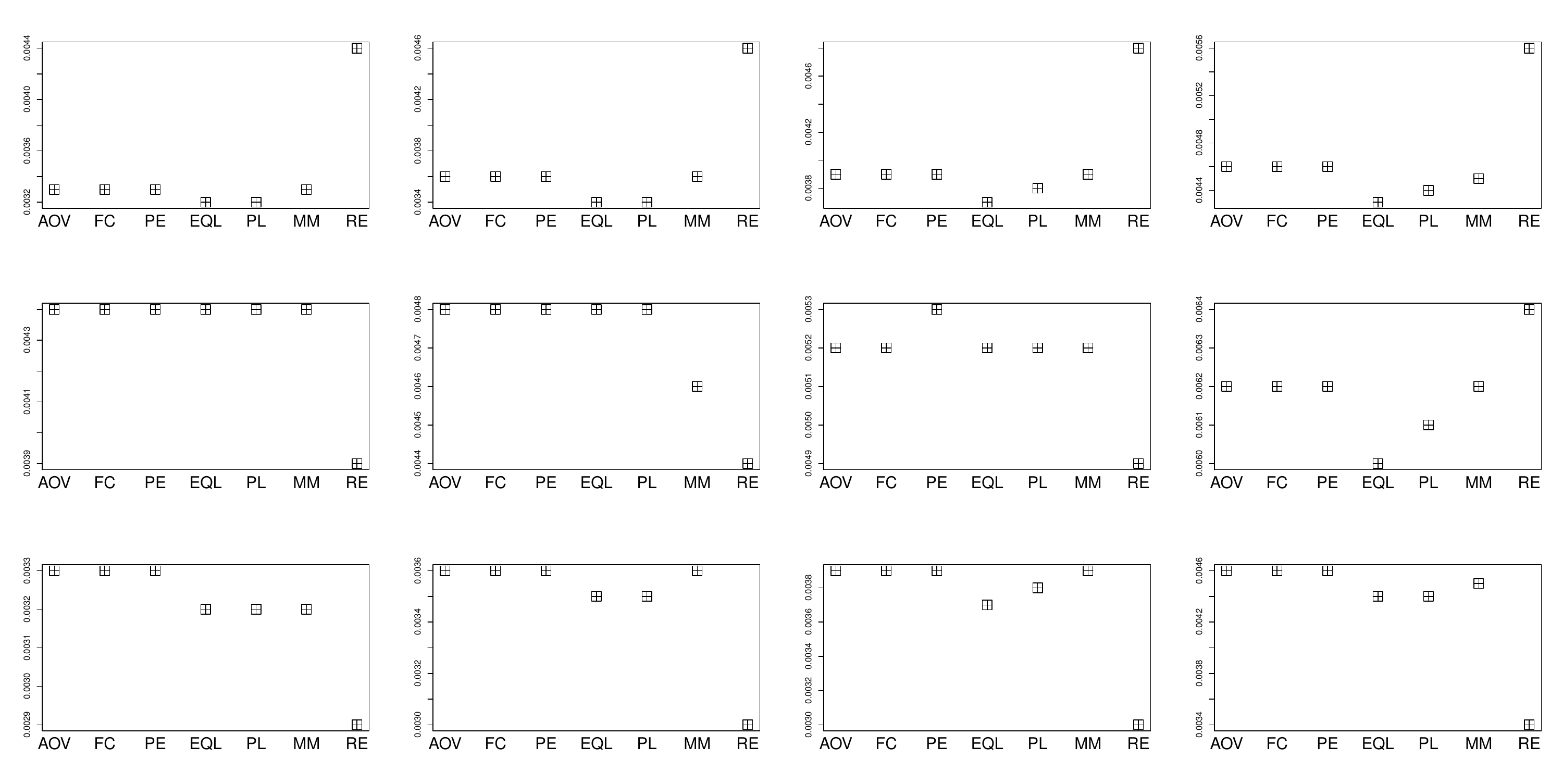}
\caption{Plot of the median MSE values alone, from the MSE boxplots given in figure 2.}
\end{center}
\end{figure}

\subsection*{Appendix C. Functional form for $\frac{\partial h}{\partial \rho}$}

\setcounter{equation}{0}
\renewcommand{\theequation}{C.\arabic{equation}}
\setcounter{figure}{0}
\renewcommand{\thefigure}{C.\arabic{figure}}

\[ h = f/g, \; \mathrm{where}\; f = \sum_{j=1}^{J}c_{j}X_{j}, \; g = \sum_{j=1}^{J}c_{j}n_{j}, c_{j} = [1 + (n_{j}-1)\rho]^{-1}. \]

\[ \frac{\partial h}{\partial \rho} = \frac{g \frac{\partial f}{\partial \rho} - f \frac{\partial g}{\partial \rho}}{g^{2}} \]

where \[ \frac{\partial f}{\partial \rho} = -\sum_{j=1}^{J} \left(\frac{X_{j}(n_{j}-1)}{[1 + (n_{j}-1)\rho]^{2}} \right)\; \mathrm{and}\;\frac{\partial g}{\partial \rho} =
-\sum_{j=1}^{J} \left(\frac{n_{j}(n_{j}-1)}{[1 + (n_{j}-1)\rho]^{2}} \right)  \]

\begin{figure}[H]
\begin{center}
\includegraphics[height=5in,width=5in,angle=0]{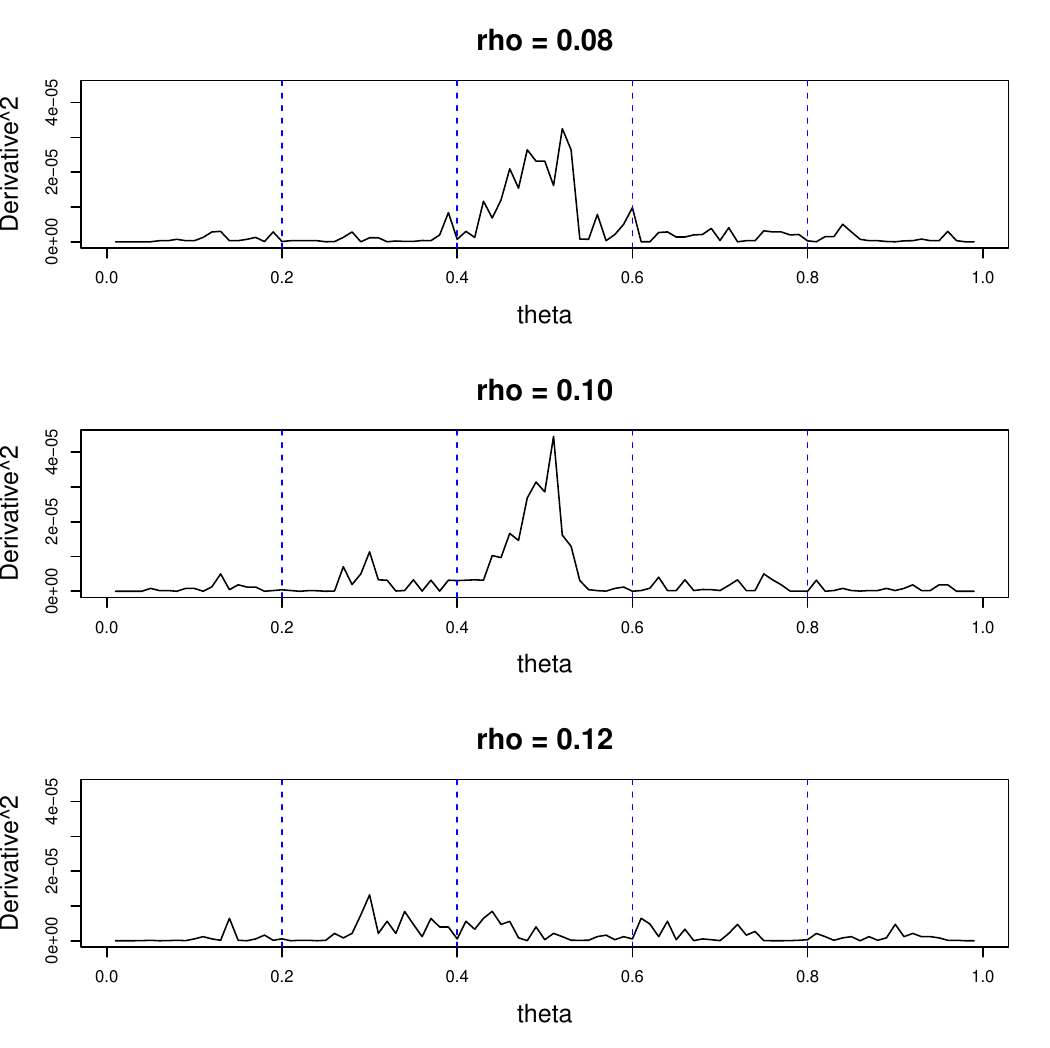}
\caption{($\frac{\partial h}{\partial \rho}$)$^{2}$ plotted as a function of $\theta$ for three different $\rho$ values. These plots were obtained by averaging across 500
simulations.}
\end{center}
\end{figure}

\subsection*{Appendix D. Histograms for $\hat{\theta}$ estimates based on method 2, for the cases with true $\theta$ = 0.25 and 0.50. }

\setcounter{equation}{0}
\renewcommand{\theequation}{D.\arabic{equation}}
\setcounter{figure}{0}
\renewcommand{\thefigure}{D.\arabic{figure}}

\newpage

\begin{figure}[H]
\begin{center}
\subfloat{\includegraphics[height=1.5in, width = 1.5in]{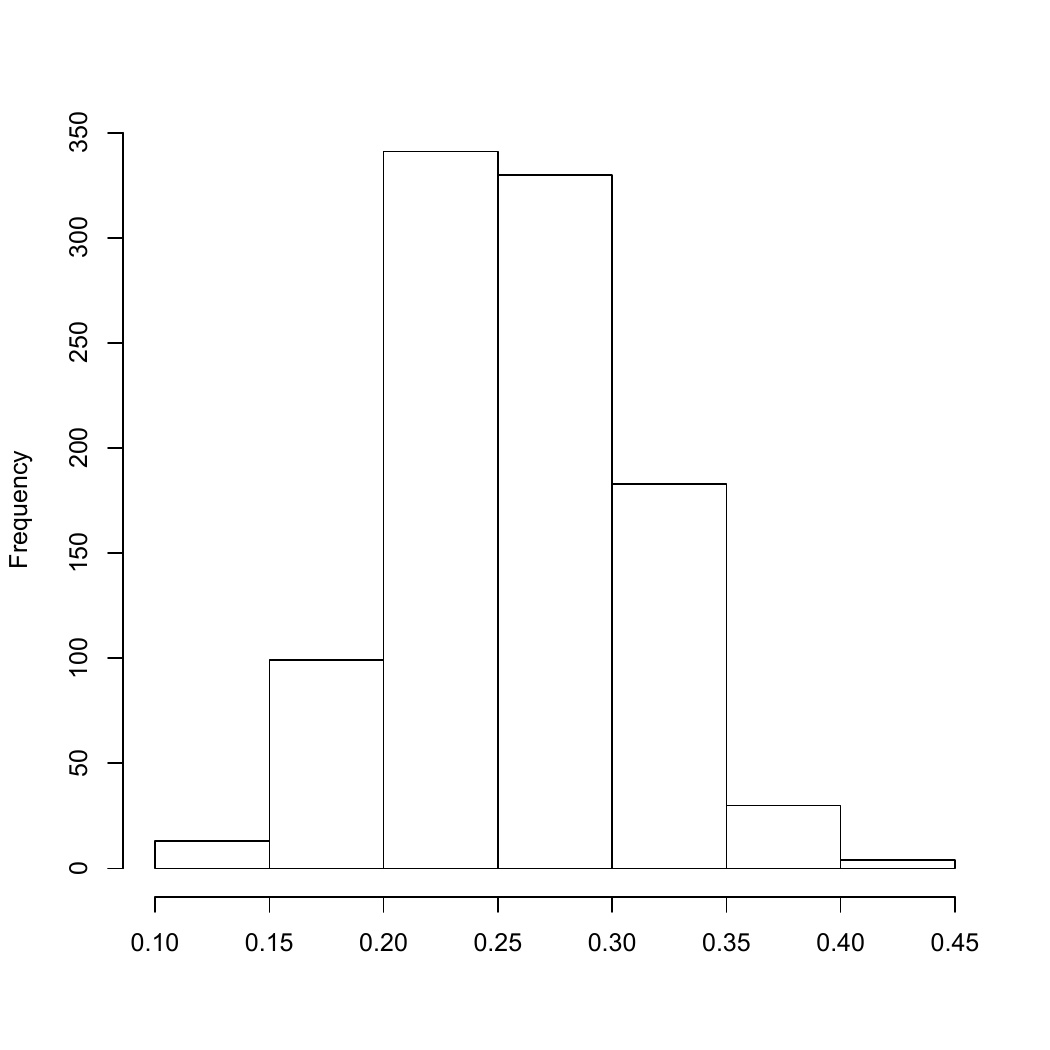}}
\subfloat{\includegraphics[height=1.5in, width = 1.5in]{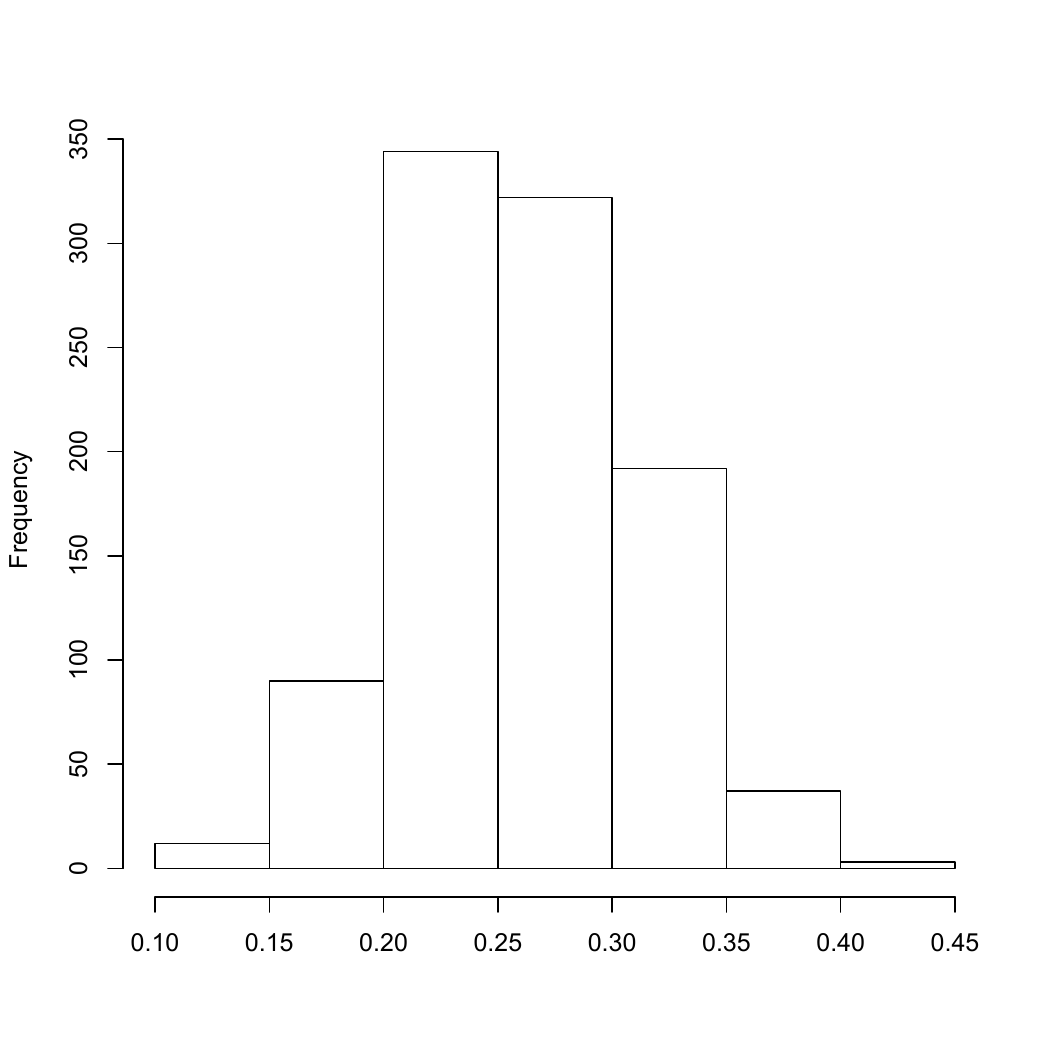}} \\
\subfloat{\includegraphics[height=1.5in, width = 1.5in]{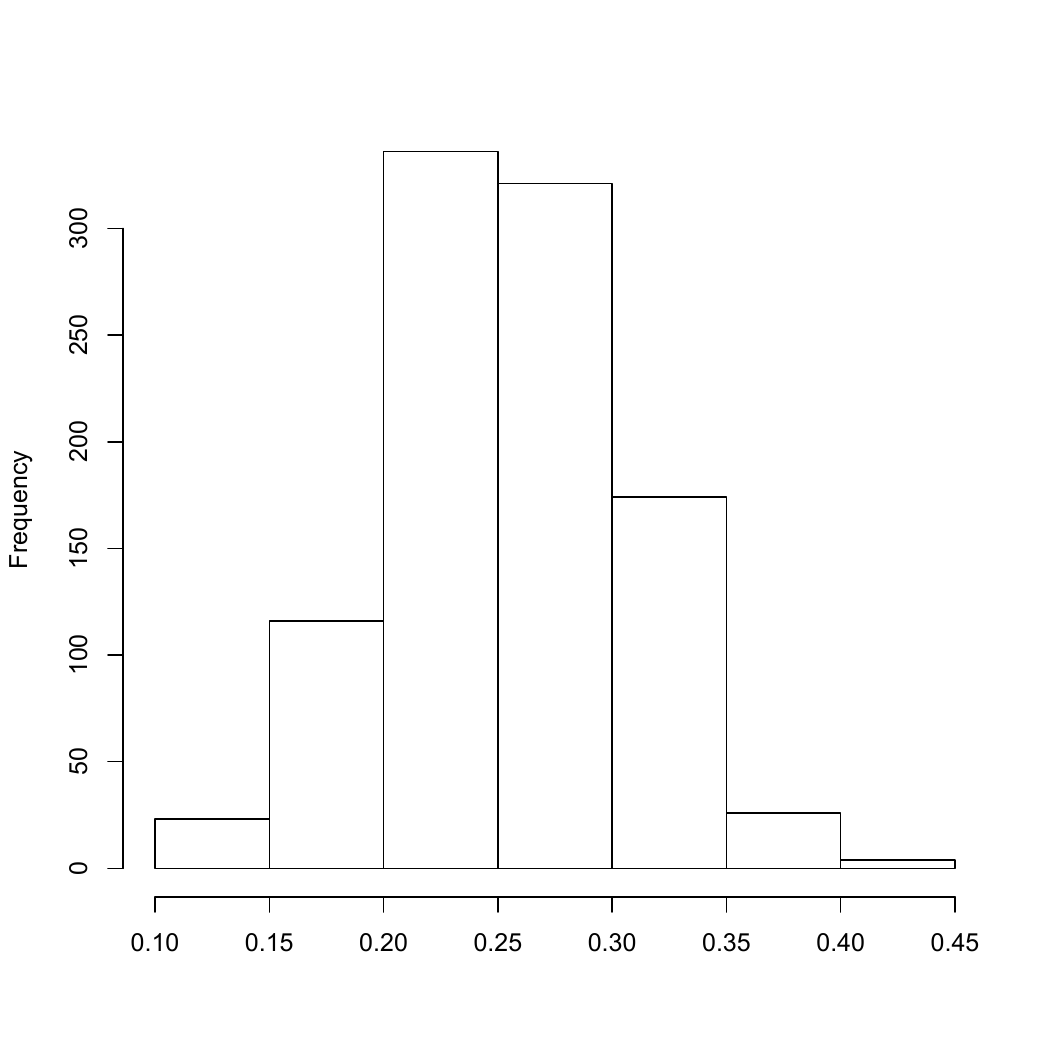}}
\subfloat{\includegraphics[height=1.5in, width = 1.5in]{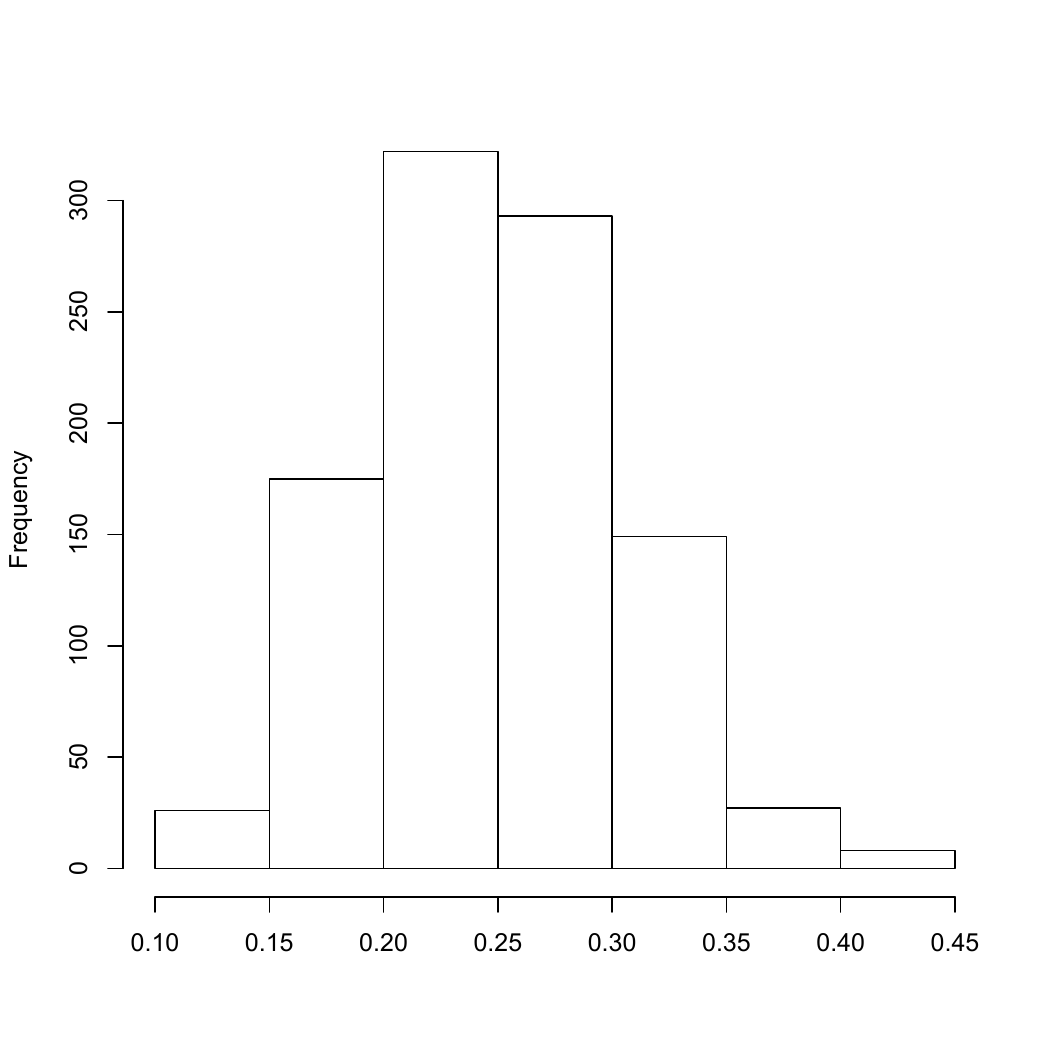}} \\
\caption{Histograms of $\hat{\theta}$ estimates based on method 2, for the cases with true $\theta$ = 0.25; clockwise from topleft, the panels correspond to $\rho$ =
0.01, 0.05, 0.10 and 0.20.}
\end{center}
\end{figure}

\begin{figure}[H]
\begin{center}
\subfloat{\includegraphics[height=1.5in, width = 1.5in]{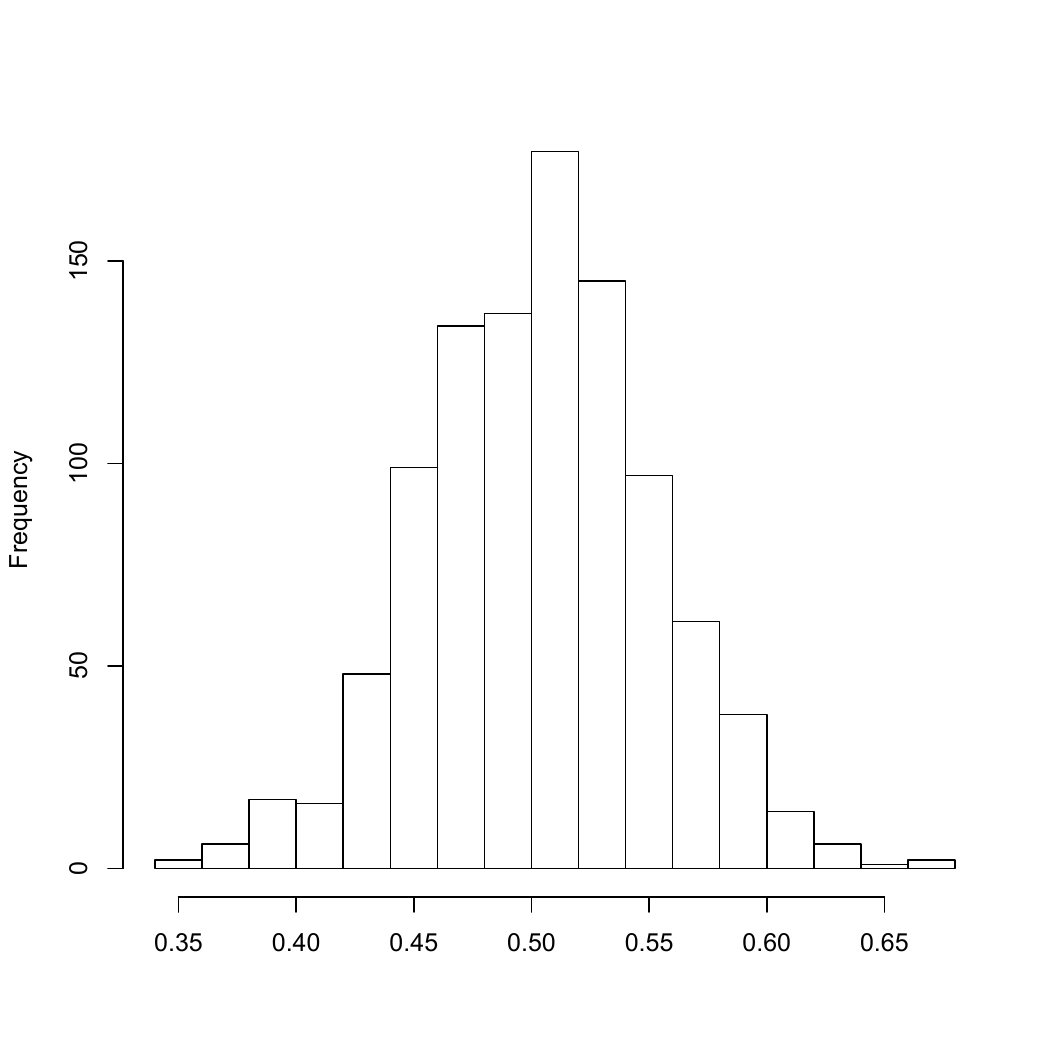}}
\subfloat{\includegraphics[height=1.5in, width = 1.5in]{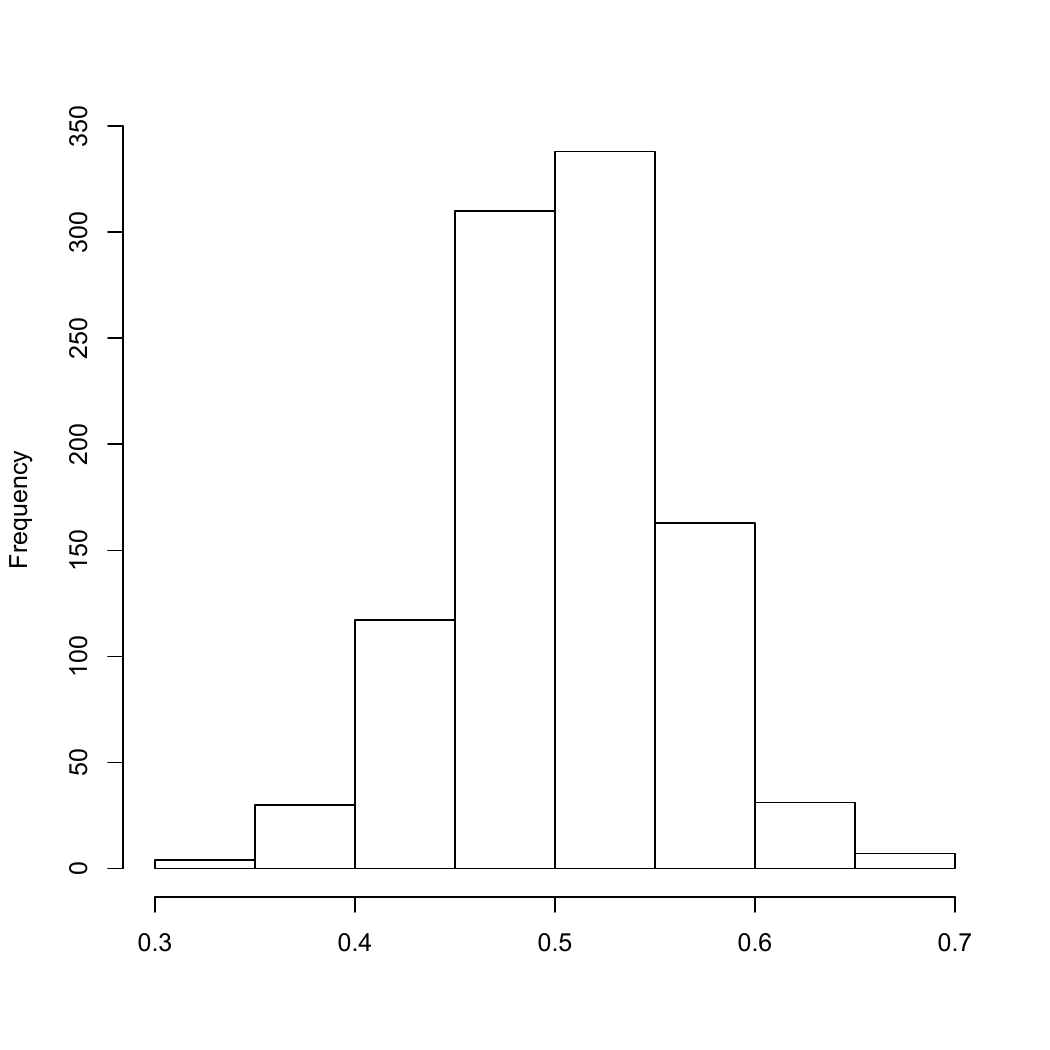}} \\
\subfloat{\includegraphics[height=1.5in, width = 1.5in]{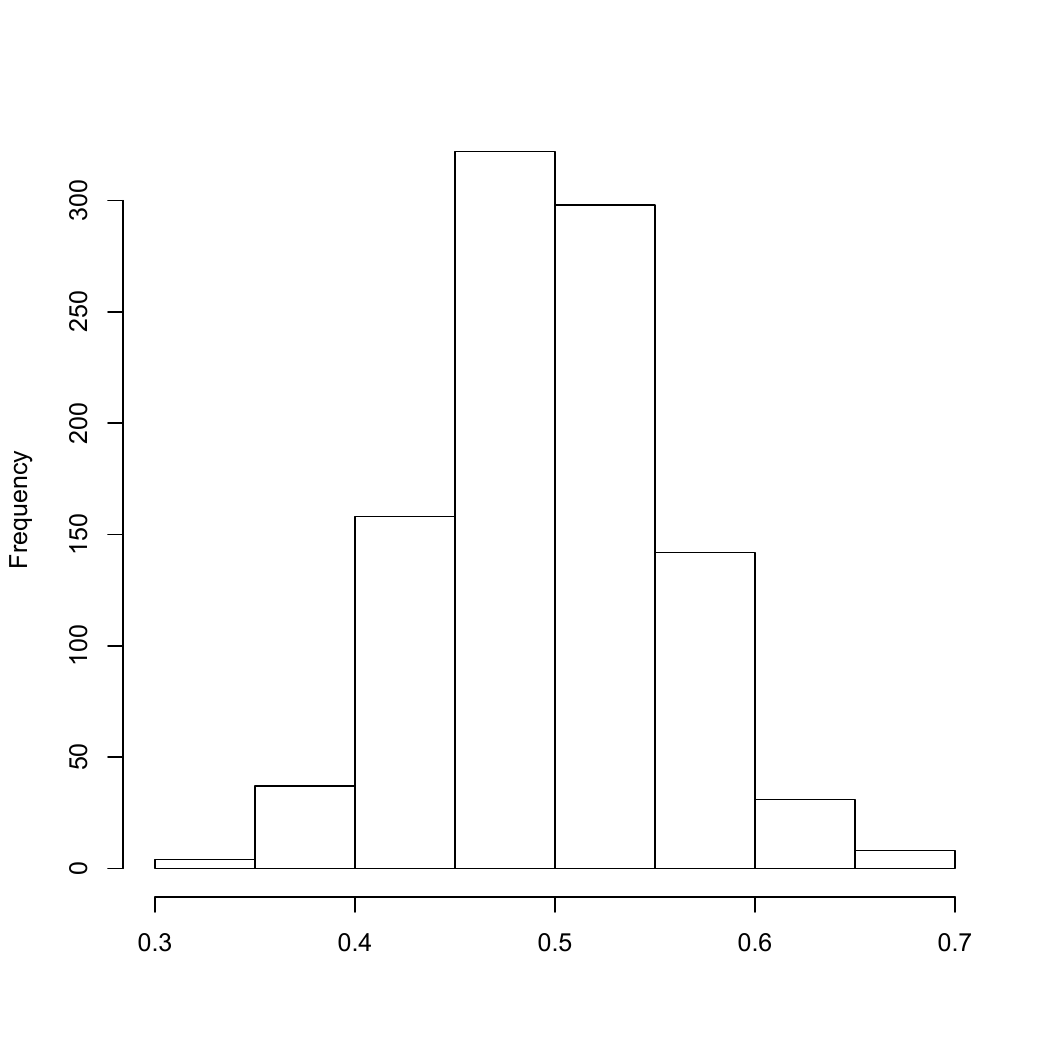}}
\subfloat{\includegraphics[height=1.5in, width = 1.5in]{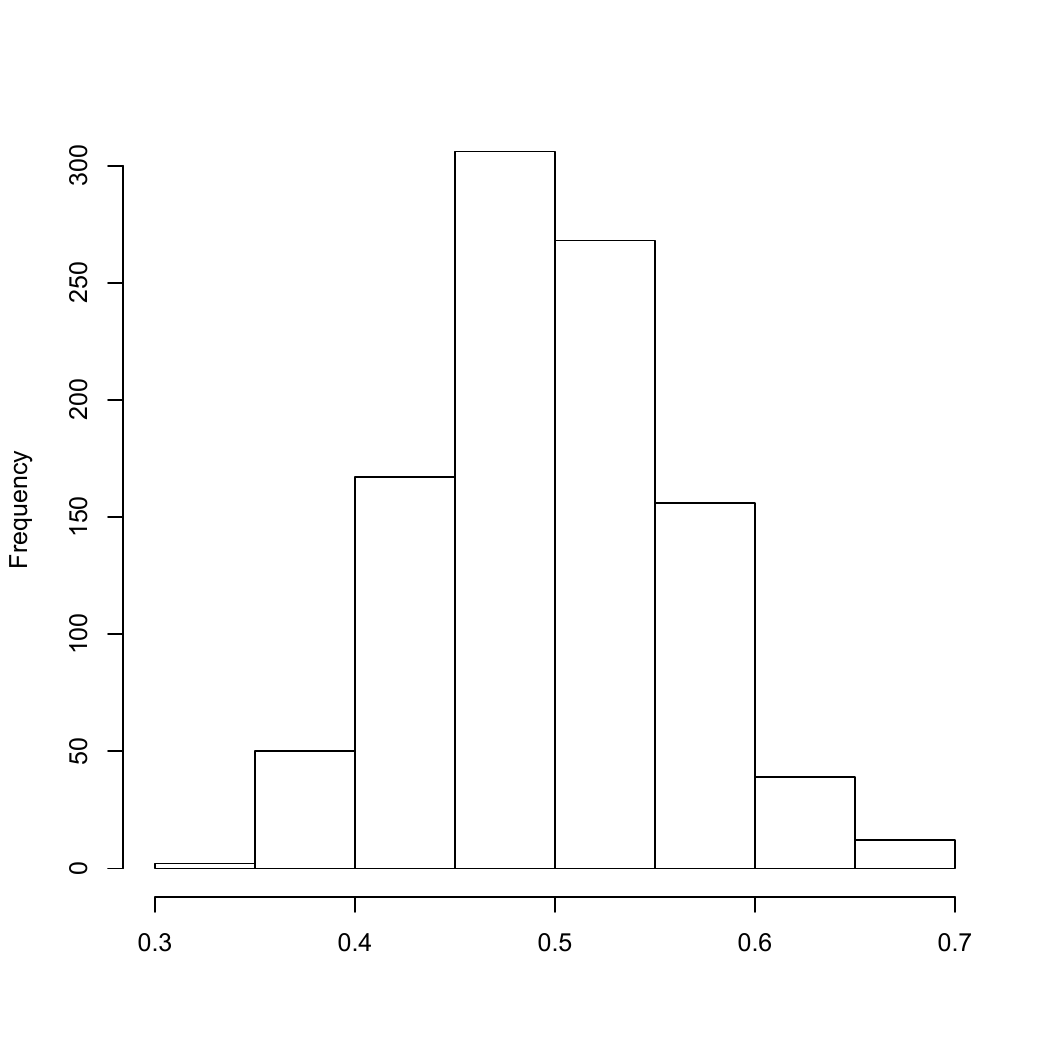}} \\
\caption{Histograms of $\hat{\theta}$ estimates based on method 2, for the cases with true $\theta$ = 0.50; clockwise from topleft, the panels correspond to $\rho$ =
0.01, 0.05, 0.10 and 0.20.}
\end{center}
\end{figure}

\subsection*{Appendix E. R codes}
\scriptsize

\begin{verbatim}
### rho estimation: AOV

rho.aov.func <- function(x.j, n.j){
                     N <- sum(n.j)
                     k <- length(n.j)
                   n.A <- (1/(k-1))*(N - (sum(n.j^2)/N))
                   msb <- (1/(k-1))*( sum((x.j^2)/n.j) - (((sum(x.j))^2)/N) )
                   msw <- (1/(N-k))*( sum(x.j) - sum((x.j^2)/n.j) )
               rho.est <- (msb - msw)/(msb + ((n.A-1)*msw))
                   return(rho.est) }

### rho estimation: FC

rho.fc.func <- function(x.j, n.j){
                     N <- sum(n.j)
                     k <- length(n.j)
                pi.hat <- sum(x.j)/sum(n.j)
               rho.est <- 1 - ( (sum((x.j*(n.j-x.j))/n.j))/((N-k)*pi.hat*(1-pi.hat)) )
                 return(rho.est) }

### rho estimation: PE

rho.pe.func <- function(x.j, n.j){
                mu.hat <- sum(x.j*(n.j-1))/sum(n.j*(n.j-1))
               rho.est <- (1/(mu.hat*(1-mu.hat)))*( ((sum(x.j*(x.j-1)))/(sum(n.j*(n.j-1)))) - ((mu.hat^2)) )
                return(rho.est) }

## rho estimation: EQL

rho.eql.func <- function(theta.est, x.j, n.j){
     eps <- 0.00000001;     x.j[x.j == 0] <- eps; x.j[x.j == n.j] <- n.j[x.j == n.j] - eps

   ff.func <- function(rho, theta, x.j, n.j){
                      phi.j <- (1 + (n.j-1)*rho)
                      dev.j <- 2*( (x.j*log(x.j/(n.j*theta))) + ((n.j-x.j)*log((n.j-x.j)/(n.j-(n.j*theta)))) )
                       rslt <- sum( ((n.j-1)*(dev.j - phi.j))/(phi.j^2) )
                       return(rslt)         }
     rho.rslt <- uniroot(ff.func, c(-0.5, 1), tol = 0.00001, theta = theta.est, x.j = x.j, n.j = n.j)$root
                       return(rho.rslt)     }

## rho estimation: PL

rho.pl.func <- function(theta.est, x.j, n.j){

   ff.func <- function(rho, theta, x.j, n.j){
                      phi.j <- (1 + (n.j-1)*rho)
                    chisq.j <- ((x.j - (n.j*theta))^2)/(n.j*theta*(1-theta))
                       rslt <- sum( ((n.j-1)*(chisq.j - phi.j))/(phi.j^2) )
                       return(rslt)         }
     rho.rslt <- uniroot(ff.func, c(-0.5, 1), tol = 0.00001, theta = theta.est, x.j = x.j, n.j = n.j)$root
                       return(rho.rslt)     }


## rho estimation: MM

rho.mm.func <- function(rho.grid, theta.est, x.j, n.j){

     ssr.rho.temp <- array(,length(rho.grid))

         for(r.ind in 1:length(rho.grid)){
                       rho <- rho.grid[r.ind]; theta <- theta.est;
                     phi.j <- (1 + (n.j-1)*rho)
ssr.j <- sum(((x.j - (n.j*theta))^2)/(n.j*(1 +(n.j-1)*rho)*theta*(1-theta)))
       ssr.rho.temp[r.ind] <- ssr.j - length(x.j) }
            ssr.rho.up.min <- min(ssr.rho.temp[ssr.rho.temp > 0])
            ssr.rho.dn.max <- max(ssr.rho.temp[ssr.rho.temp < 0])
                   rho.est <- mean(c(rho.grid[ssr.rho.temp == ssr.rho.up.min], rho.grid[ssr.rho.temp == ssr.rho.dn.max]))
                                       return(rho.est) }

## theta estimation, for rho based on AOV, FC, or PE

       x.j <- ## array of x.j data
       n.j <- ## array of n.j data
       J <- length(x.j)

         rho.est <- rho.aov.func(x.j, n.j) # or rho.fe.func(x.j, n.j) or rho.pe.func(x.j, n.j)
             c.j <- 1/(1 + (n.j-1)*rho.est)
       theta.est <- sum(c.j*x.j)/sum(c.j*n.j)

## jackknife based estimates may be obtained as follows

       jk.len <- length(x.j); theta.jk.vec <- rho.jk.vec <- array(,jk.len)

       for(jk.ind in 1:jk.len){
                      x.j.jk <- x.j[-jk.ind]; n.j.jk <- n.j[-jk.ind]
                      rho.jk <- rho.aov.func(x.j.jk, n.j.jk)  # or rho.fe.func(x.j.jk, n.j.jk) or rho.pe.func(x.j.jk, n.j.jk)
          rho.jk.vec[jk.ind] <- rho.jk
                      c.j.jk <- 1/(1 + (n.j.jk-1)*rho.jk)
                    theta.jk <- sum(c.j.jk*x.j.jk)/sum(c.j.jk*n.j.jk); theta.jk.vec[jk.ind] <- theta.jk
                              }
               theta.jk.var <- sum((theta.jk.vec - theta.est)^2)
                 rho.jk.var <- sum((rho.jk.vec - rho.est)^2)

# theta estimation, for rho based on EQL, PL or MM (where theta and rho estimation alternates until convergence)
  max.n.iter <- # e.g. 200
    for(m in 1:max.n.iter){

      if(m == 1){
           rho <- 0
           c.j <- 1/(1 + (n.j-1)*rho)
           theta.est <- sum(c.j*x.j)/sum(c.j*n.j)
           # rho.grid <- seq(-0.2, 1.0, 0.0001) ## uncomment this for MM
rho.est <- rho.eql.func(theta.est, x.j, n.j)# or rho.pl.func(theta.est, x.j, n.j) or rho.mm.func(rho.grid, theta.est, x.j, n.j)
         diff.max <- 10
                }
      if(m > 1){
           rho <- rho.est; theta <- theta.est
           c.j <- 1/(1 + (n.j-1)*rho)
            theta.est <- sum(c.j*x.j)/sum(c.j*n.j)
           theta.diff <- abs(theta.est - theta)
           #  rho.grid <- seq(-0.2, 1.0, 0.0001) ## uncomment this for MM
          rho.est.old <- rho.est
rho.est <- rho.eql.func(theta.est, x.j, n.j)# or rho.pl.func(theta.est, x.j, n.j) or rho.mm.func(rho.grid, theta.est, x.j, n.j)
         rho.abs.diff <- abs(rho.est - rho.est.old)
             diff.max <- max(c(rho.abs.diff, theta.diff))
         if(diff.max < 0.000005){break} }
                  }

\end{verbatim}

This work was completed before the start of the COVID-19 pandemic in the USA, and the references have not been updated after February, 2020.

\end{document}